# Making the Graph Laplacian Physical: Multiscale Coarse-Graining in Electrical Oscillator Networks

**Juan Bisquert**

Instituto de Tecnología Química (ITQ), Consejo Superior de Investigaciones Científicas-Universitat Politècnica de València, Valencia, Spain.

Corresponding author email: jbisquer@itq.upv.es

## Abstract

Spectral gaps are widely interpreted as signatures of collective organization, yet it is rarely clear whether the resulting modes correspond to physical variables or merely to convenient mathematical coordinates. Here we show that, in resistor-coupled LC networks, the graph Laplacian defines an experimentally accessible hierarchy of electrical descriptions. Its eigenvectors become measurable voltage patterns, while its eigenvalues quantify the coupling-induced resistive damping of those patterns. When coupling is strong within regions and weak between them, microscopic voltages collapse into nested collective variables. A network of 1002 resonators reduces first to 42 regional voltages and then to four geometry-controlled global modes. Summing the microscopic Kirchhoff equations produces a directly constructible 42-node RLC circuit, without parameter fitting, that reproduces regional dynamics with 1.4–4.6% global RMS error. Finer modal coordinates improve accuracy but no longer correspond generally to simple resistor graphs. Robustness tests show that the regional level survives substantial conductance disorder and internal rewiring, but weakens when perturbations impair mixing within a region; changes in interregional connectivity selectively disrupt the global level. Thus the Laplacian spectrum becomes a physical design principle: it identifies which collective electrical variables emerge, how they can be constructed, and when they provide an adequate reduced description.

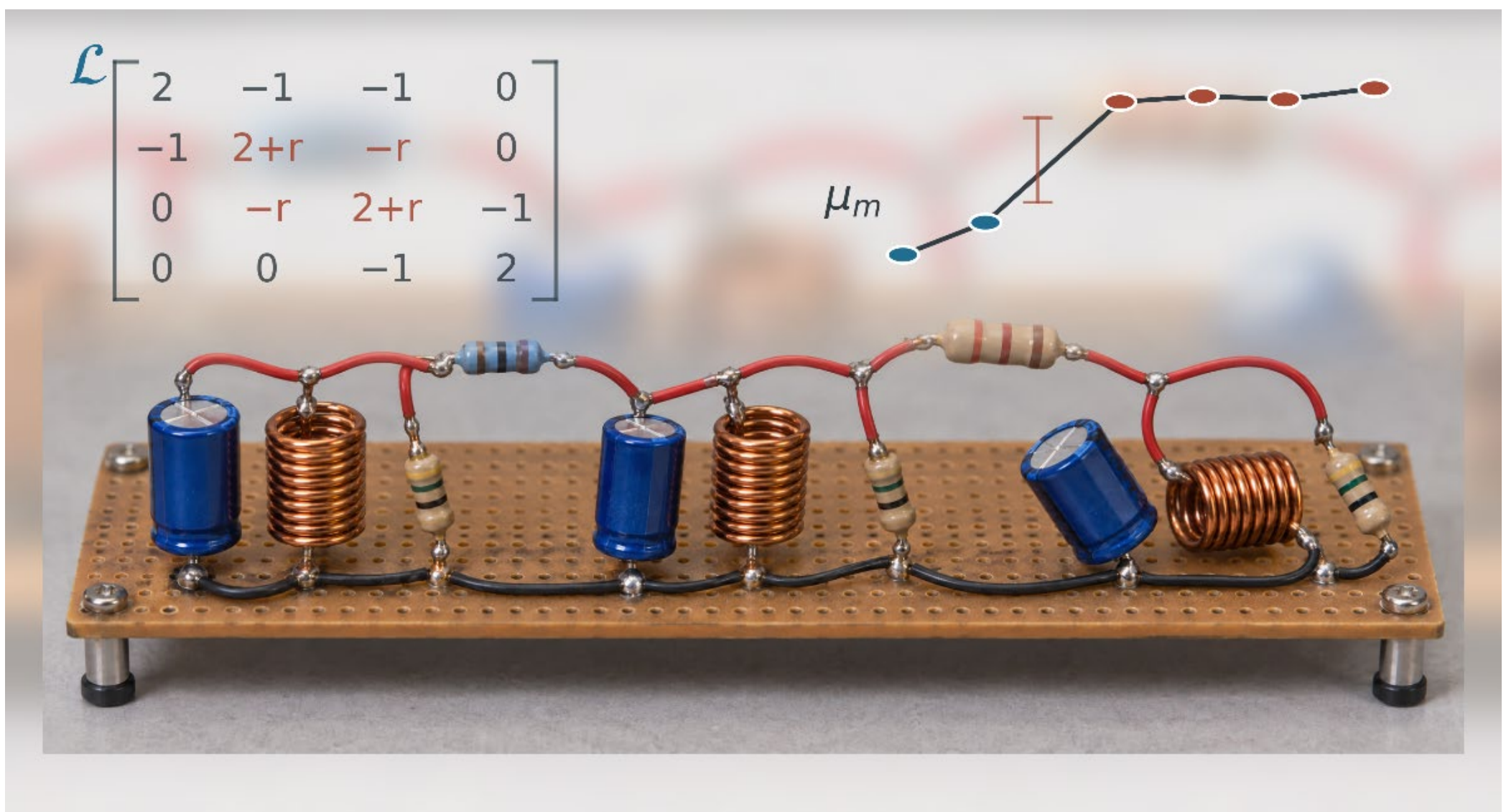

# 1. Introduction

Many complex systems contain far more microscopic degrees of freedom than are required to describe their observable dynamics. The central problem is therefore not only to solve a large network, but to identify the smaller set of variables that carries its collective response. This question appears in electrical and mechanical networks, coupled oscillators, diffusion, power grids, network control and biological systems [1–8]. In all these settings, connectivity is encoded naturally by a graph, and the graph Laplacian provides the operator that penalizes differences between connected nodes [9–12].

The Laplacian spectrum is widely used to characterize connectivity, mixing, synchronization and community structure [9–16]. Low eigenvectors vary gradually over strongly connected portions of a graph, whereas higher eigenvectors contain increasingly local contrasts. Spectral gaps can therefore separate families of patterns. This principle underlies spectral clustering, graph partitioning and modern graph-reduction algorithms [17–22]. However, two conceptual difficulties remain. First, the eigenvectors are often treated as mathematical coordinates without a direct realization. Second, a spectrum alone does not establish that a reduced model reproduces the time evolution or transfer response of the original physical system.

Recent proposals in which isolated spectral states acquire an emergent information-theoretic meaning provide an additional motivation [23]. Our question is narrower and physical: for a specified passive electrical network, what do the spectrum and its separated subspaces correspond to in measurable circuit variables? We therefore work with the weighted Laplacian, retain the complete component-level dynamics, and test every proposed reduction against the response of the original circuit.

Oscillator networks provide an important context. Work on phase synchronization has established deep relations between topology, Laplacian eigenvalues and collective dynamics [3–7,13–16]. The present system is deliberately different: its nodes are passive linear LC resonators, not self-sustained phase oscillators, and no claim of Kuramoto synchronization is made. Resistive coupling causes voltage differences to relax and selects collective resonant patterns. This restriction is useful because every term in the model has an ordinary circuit interpretation and the complete network can, in principle, be assembled from capacitors, inductors and resistors.

A related motivation comes from brain-network research. Neural populations exhibit low-dimensional collective activity despite vast microscopic complexity, and connectome eigenmodes have been used to relate anatomical structure to functional patterns [24–32]. Our electrical network is not a neuronal model: it contains no spikes, thresholds, synapses, plasticity or self-generated oscillations. The connection is instead methodological. Both problems ask how connectivity constrains observable collective coordinates and how many variables are needed at a given temporal and spatial resolution.

Here we establish a physical coarse-graining principle for resistor-coupled resonator networks. We first introduce a 252-node spherical circuit in which the meaning of the Laplacian spectrum can be seen directly. We then examine a 1002-node network partitioned into 42 regions, identify the hierarchy 1002→42→4, and test its robustness against five distinct connectivity perturbations. Finally, we construct reduced electrical models and compare their free responses with those of the complete circuit. The combined results support a scale-selective statement: each emergent level is insensitive to many changes below that level, but responds strongly to changes in connectivity at its own scale.

# 2. Electrical network model

## 2.1 Nodal equations and weighted Laplacian

Each node i contains a parallel capacitor C, inductor L and leakage conductance $G_0$. Nodes i and j are connected by a resistor $R_{ij}$ with conductance $G_{ij} = \frac{1}{R_{ij}}$. Kirchhoff's current law gives

$$C\frac{dv_i}{dt} + G_0 v_i + \left(\frac{1}{L}\right)\int v_i dt + \sum_j G_{ij}\left(v_i - v_j\right) = I_i^{ext}(t). \tag{1}$$

Collecting all node voltages in $v$ and all applied currents in $I_{ext}$ yields

$$C\frac{dv}{dt} + G_0 v + \left(\frac{1}{L}\right)\int v dt + G_c\mathcal{L}v = I^{ext}(t), \tag{2}$$

where $\mathcal{L} = D - A$ is the weighted graph Laplacian after extracting a reference coupling conductance $G_c$. Its diagonal entries are $\mathcal{L}_{ii} = \sum_j w_{ij}$ and its off-diagonal entries are $\mathcal{L}_{ij} = -w_{ij}$. The dimensionless weight is $w_{ij} = \frac{G_{ij}}{G_c}$. Because the coupling resistors are passive and reciprocal, $\mathcal{L}$ is real, symmetric and positive semidefinite. The uniform vector is its zero eigenvector for a connected graph.

The network is embedded on a triangulated sphere (Fig. 1). Nearly uniform node positions are generated by recursively subdividing an icosahedron and projecting the new vertices onto the unit sphere. Each node is connected to its geometric nearest neighbours by the triangulation. Regions are defined by assigning every node to its closest reference direction. Edges within the same region have resistance $R_s$; edges crossing a regional boundary have resistance $R_L$. The control parameter is

$$r = \frac{G_L}{G_s} = \frac{R_s}{R_L}. \tag{3}$$

Small $r$ therefore means strong internal coupling relative to interregional coupling. Unless otherwise indicated, the calculations use r = 0.03. The absolute values of C, L and conductance set the time and current scales; the modal organization depends on the dimensionless Laplacian and on $r$.

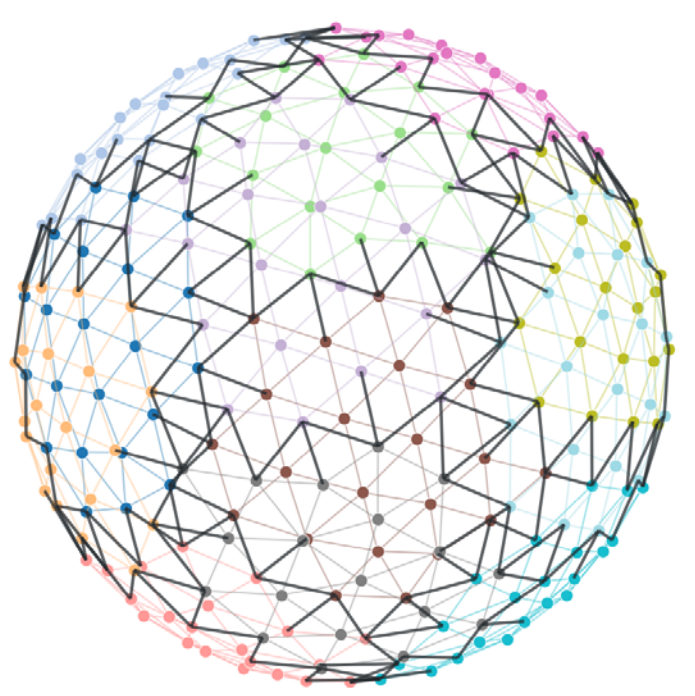

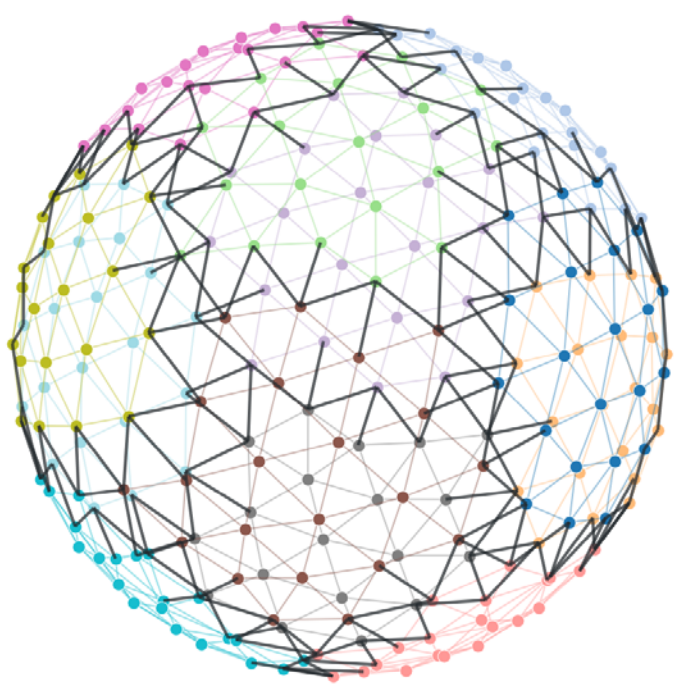

**Figure 1. Physical spherical network used to introduce the model.** The 252 identical LC resonators are distributed nearly uniformly over a triangulated sphere and divided into 12 regions. Coloured connections carry the strong internal conductance $G_s$, while dark boundary connections carry the weaker conductance $G_L$. The two views show that the regional partition covers the complete closed surface without privileged boundaries.

## 2.2 Electrical meaning of eigenvalues and eigenmodes

Let $\varphi_m$ be a normalized eigenvector of the Laplacian with eigenvalue $\mu_m$,

$$\mathcal{L}\varphi_m = \mu_m \varphi_m, \quad 0 = \mu_1 \le \mu_2 \le \cdots \le \mu_N. \tag{4}$$

Expanding $v(t) = \sum_m q_m(t)\varphi_m$ diagonalizes the coupling term. Every modal coordinate obeys an independent resonator equation,

$$C\frac{dq_m}{dt} + (G_0 + G_c\mu_m)q_m + \left(\frac{1}{L}\right)\int q_m dt = I_m(t), \tag{5}$$

with $I_m = \varphi_m^{\mathrm{T}} I^{\mathrm{ext}}$. The eigenvector is therefore a physically measurable voltage pattern, while $G_c\mu_m$ is the additional coupling conductance associated with that pattern and determines its coupling-induced damping. Small $\mu_m$ indicates a pattern with little voltage difference across connected nodes and hence weak resistive damping. Large $\mu_m$ identifies a pattern containing strong local voltage contrasts and greater dissipation. An input pattern excites mode $m$ in proportion to its overlap $\varphi_m^{\mathrm{T}} I^{\mathrm{ext}}$.

A spectral gap separates two families of voltage patterns according to the conductance penalty imposed by the coupling network. If $\mu_K \ll \mu_{K+1}$, the first $K$ modes experience much less coupling-induced damping than the higher modes and therefore define a candidate $K$-dimensional collective subspace. The gap suggests how many variables may be retained, but it does not determine what those variables physically represent. This information comes from the eigenvectors. If the low eigenvectors are nearly uniform within strongly connected regions, their subspace can be represented by one measurable voltage per region. Thus, both spectral separation and spatially interpretable eigenvectors are required to justify a reduced regional circuit.

A minimal example makes this concrete before turning to the large spherical networks. Consider six nodes forming two triangles, $\{1,2,3\}$ and $\{4,5,6\}$, each internally complete with unit edge weight, joined by a single weak bridge edge (3,4) of weight $r$. Ordering the nodes 1–6, its weighted Laplacian is

$$\mathcal{L} = \begin{bmatrix} 2 & -1 & -1 & 0 & 0 & 0 \\ -1 & 2 & -1 & 0 & 0 & 0 \\ -1 & -1 & 2+r & -r & 0 & 0 \\ 0 & 0 & -r & 2+r & -1 & -1 \\ 0 & 0 & 0 & -1 & 2 & -1 \\ 0 & 0 & 0 & -1 & -1 & 2 \end{bmatrix}$$

For a weak bridge, $r = 0.1$, direct diagonalization gives the eigenvalues $\mu = 0, 0.064, 3, 3, 3, 3.136$.

The zero eigenvalue is the uniform mode $\varphi_1$, present for any connected graph. The next eigenvalue is separated from the remaining four by nearly two orders of magnitude. The resulting two-dimensional low subspace comprises the uniform mode and one nonzero collective mode, separated by a spectral gap from the four internal modes. This is the same two-level organization that structures the much larger networks studied below [23]. The nonzero low eigenvector is close to a positive constant on nodes $\{1,2,3\}$ and a negative constant on nodes $\{4,5,6\}$ (numerically approximately $0.42, 0.42, 0.39, -0.39, -0.42, -0.42$): this mode measures the voltage difference between the two triangles and is weakly damped because it involves only the single weak bridge conductance.

The remaining four eigenvalues cluster near $\mu = 3$, because each isolated triangle $K_3$ contributes two degenerate internal modes with eigenvalue 3. The weak bridge leaves three of these eigenvalues unchanged and shifts the fourth to 3.136. This six-node toy network already exhibits, in miniature, the two-level organization that the 252- and 1002-node spherical networks exhibit at a much larger scale.

### 2.3 Evolution under electrical stimulation

For the time-domain calculations we set $C = L = G_c = 1$ and $G_0 = 0.08$ in normalized units. The capacitor voltage is initialized either at one node, uniformly over one region, or at two separated nodes, while all inductor currents initially vanish. After this preparation the network undergoes a free response with $I^{ext} = 0$. Because the system is linear, the evolution is calculated exactly in the Laplacian eigenbasis, without time-stepping error. The normalized parameters display modal decay on a common timescale; the dimensioned reference values ($R_s = 100$ Ω, $C = 10$ μF, $L = 10$ mH and $R_0 = 5$ kΩ) serve only to demonstrate practical component scales (Supporting Information S1). Differentiating Eq. (5) with respect to time removes the integral term; with $I^{ext} = 0$ for t > 0 this gives

$$\frac{d^2 q_m}{dt^2} + (G_0 + \mu_m)\frac{dq_m}{dt} + q_m = 0. \quad (6)$$

The observed propagation is not a movement of a single object along the surface. It is the interference of many spatial eigenmodes with different damping. Early times contain local and internal modes; later responses are progressively dominated by the low collective spectrum. Uniform regional stimulation overlaps mainly with the low subspace, whereas a point stimulus necessarily excites higher internal modes.

## 3. Collective organization of the 1002-node network

### 3.1 Forty-two regional electrical variables

The large system contains 1002 resonators and 3000 triangulation edges. Forty-two reference directions define compact regions containing approximately 22–28 nodes each (Fig. 2). We first ask whether this geometrical partition generates a distinct collective level in the electrical dynamics.

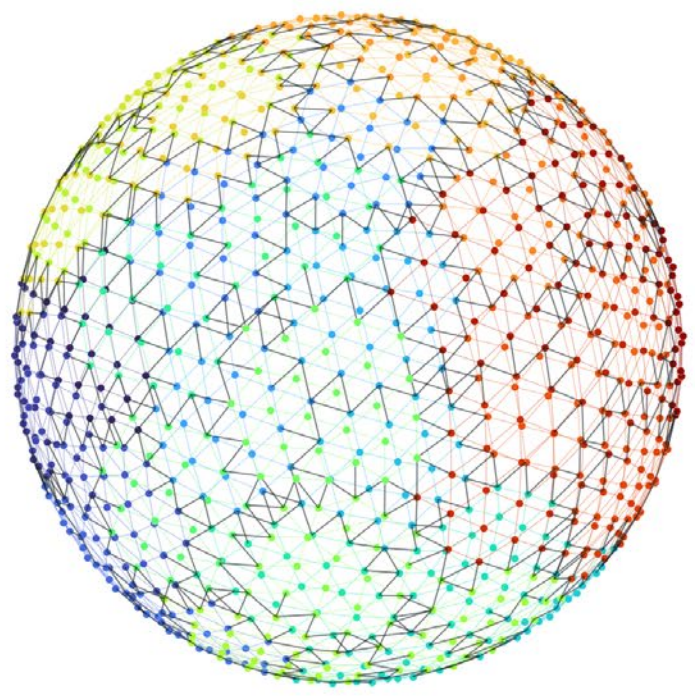

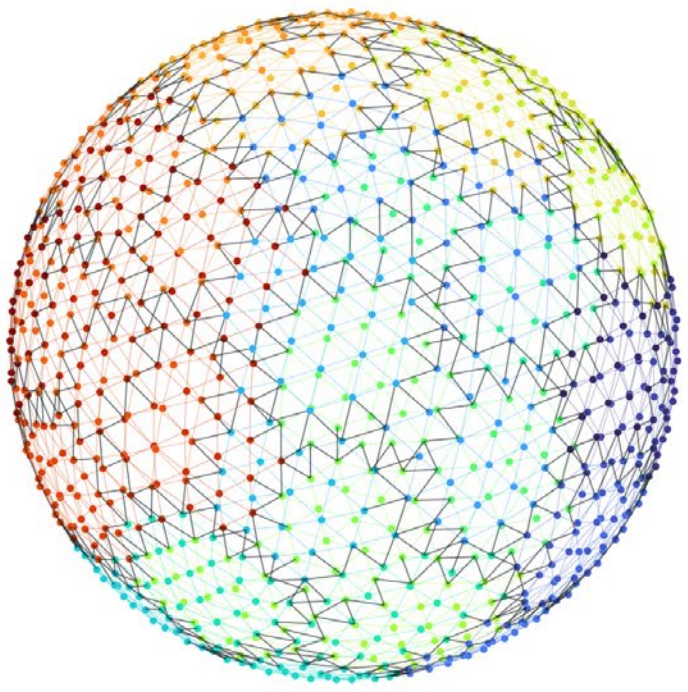

**Figure 2. Geometry of the principal 1002-node network.** The nodes are partitioned into 42 approximately equal spherical regions. Internal connections are strong and coloured by region, whereas dark connections cross regional boundaries. The network remains locally triangulated while its conductance hierarchy distinguishes internal and interregional organization.

At $r = 0.03$, the ordered Laplacian spectrum displays a well-separated low band containing exactly 42 modes (Fig. 3a). The spectral boundary lies between $\mu_{42}$, the highest eigenvalue within this band, and $\mu_{43}$, the first eigenvalue of the remaining spectrum. The separation is quantified by

$$g_{42} = \frac{\mu_{43}}{\mu_{42}} = 8.08.$$

Thus, the first mode outside the low band carries more than eight times the coupling-conductance penalty of the last mode within it. The spectrum therefore identifies 42 as a candidate reduced dimension. However, the physical meaning of these variables must be established from the corresponding eigenvectors.

As shown in Fig. 4, the first 42 eigenvectors are nearly uniform within each region and differ mainly in the relative voltages assigned to different regions. Their mean regional character, $\overline{\chi}$, is obtained by averaging $\chi_m$ over the 41 nonuniform modes $m = 2, \dots, 42$. Here, $\chi_m$ is the fraction of the variance of eigenvector $m$ reproduced by its 42 regional averages (Supporting Information S3). For the unperturbed network, $\overline{\chi} = 0.9974$. Mode $m = 42$ is the last mode retaining this predominantly regional structure, whereas mode $m = 43$ introduces substantial voltage variation within individual regions.

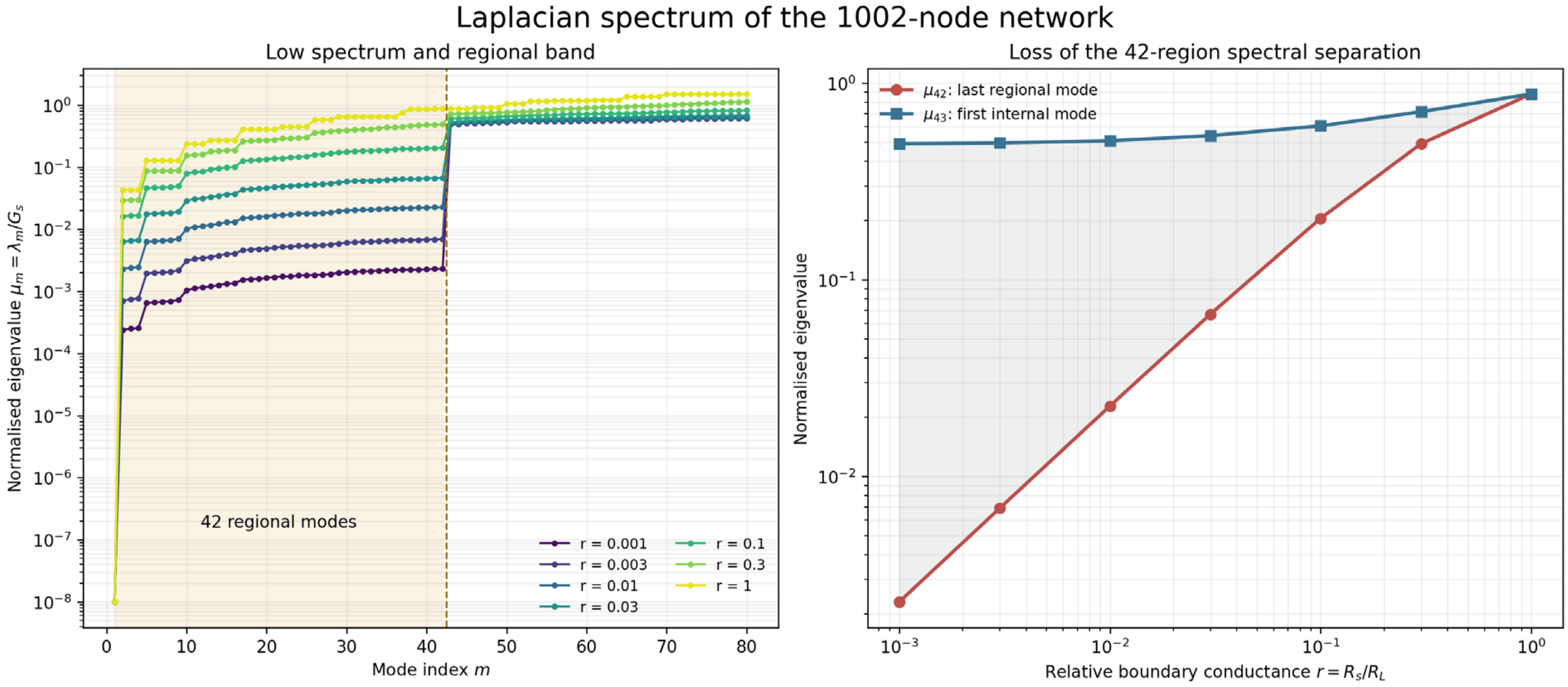

**Figure 3. Spectral classification of the 1002-node network.** (a) Low-eigenvalue spectrum for representative values of $r$, showing the band formed by the first 42 regional modes. (b) Evolution of the last regional eigenvalue, $\mu_{42}$, and the first internal eigenvalue, $\mu_{43}$. Their separation is greatest when the interregional conductance is weak and decreases progressively as $r$ approaches unity.

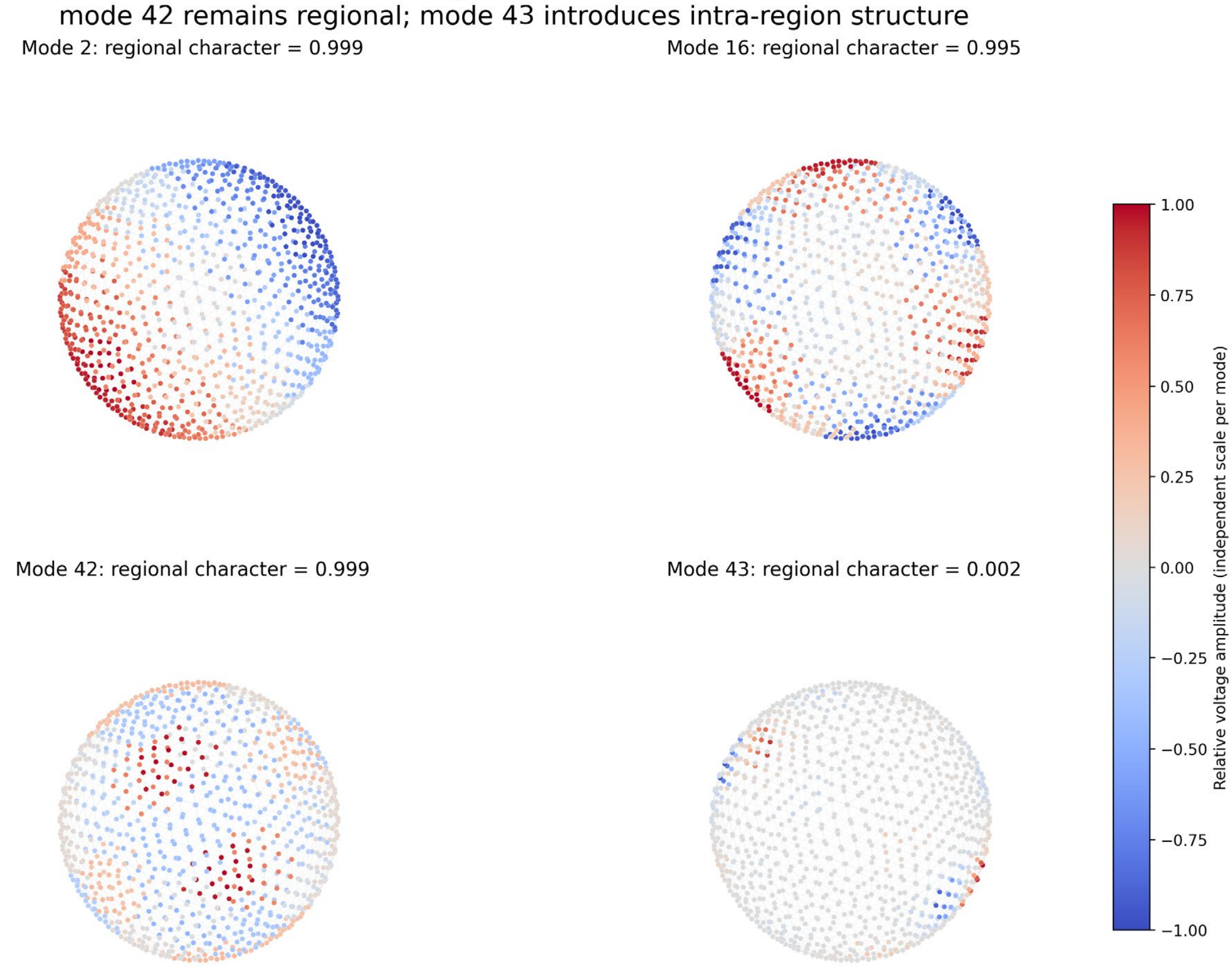

**Figure 4. Physical meaning of representative eigenvectors in the 1002-node system at $r = 0.03$.** Low and intermediate eigenvectors are nearly uniform within each region. Mode $m = 42$ is the final member of the regional band, whereas mode $m = 43$ is the first mode containing substantial intraregional voltage structure. Colour denotes signed voltage amplitude, independently normalized for each mode.

The agreement between the number of low modes and the number of regions is therefore not merely numerical: the low-dimensional spectral subspace is physically represented by one measurable voltage per region. This establishes the first effective electrical reduction,

$$1002 \text{ microscopic voltages } \rightarrow 42 \text{ regional voltages.} \quad (7)$$

### 3.2 Connectivity controls the spectral hierarchy

The regional spectral band is produced by the contrast between strong intraregional conductance and weaker interregional conductance. Changing $r$ therefore changes the separation between the regional and internal modes. For small $r$, voltage differences within each region are strongly penalized, whereas differences between regions carry a much smaller conductance penalty. The first 42 eigenvalues consequently form the narrow low band visible in Fig. 3a.

As $r$ increases, neighbouring regions become more strongly coupled. The regional eigenvalues, including $\mu_{42}$, increase because the corresponding voltage patterns produce larger currents across regional boundaries. By contrast, the first internal eigenvalue, $\mu_{43}$, changes much less because it is governed primarily by conductances within the regions. Consequently, $g_{42}$ decreases and the regional spectral gap closes progressively, as shown in Fig. 3b. The 42-variable description is therefore not an arbitrary partition imposed on the network: it exists only while equilibration within each region is substantially stronger than equilibration between regions.

The quotient network obtained by collecting the boundary conductances contains a second spectral separation after four modes. The lowest eigenvalue, ( $\nu_1 = 0$), corresponds to the uniform voltage pattern required for any connected graph. The following three eigenvalues, $\nu_1, \nu_2$ and $\nu_3$, define an approximately three-dimensional directional subspace and are separated from the next mode by the quotient gap $g_4 = \nu_5 / \nu_4$. This triplet is not generated by modularity alone. It arises because the interregional graph approximates a homogeneous two-dimensional spherical surface: its three low nonuniform eigenvectors are discrete analogues of the ($1\ell = 1$) spherical-harmonic subspace spanned by the coordinate functions (x), (y), and (z). Individual numerical eigenvectors may be rotated linear combinations of these functions because the three eigenvalues are nearly degenerate. Together with the uniform mode, they produce the four-dimensional global level whose geometrical origin is examined further in the Discussion and illustrated in Fig. 9. Consequently, the large network supports a second reduction,

$$42 \text{ regional voltages } \rightarrow 4 \text{ global voltage modes.} \quad (8)$$

Combining Eqs. 7 and 8 gives $1002 \rightarrow 42 \rightarrow 4$. The two arrows have different physical origins: strong internal conductance generates the 42-variable regional level, whereas the topology and approximate spherical symmetry of the interregional network select the four-mode global level.

### 3.3 Robustness is selective across scales

We next tested whether the hierarchy depends on the regularity of the constructed network or persists under substantial changes in microscopic connectivity. The perturbation protocols and numerical procedures are described in Supporting Information S5, and the principal results are summarized in Fig. 5. Five perturbation families were considered, with eight random realizations for every nonzero perturbation level. The perturbations consisted of lognormal disorder in the internal conductances, deletion of internal edges while preserving connected regions, degree-preserving rewiring of internal edges, weakening of one selected region, and replacement of boundary edges by long-range interregional shortcuts.

The stability of the hierarchy was evaluated using four quantities: the regional spectral gap $g_{42}$; the mean regional character $\overline{\chi}$ over modes $m = 2, \dots, 42$; the secondary quotient gap $g_4 = \nu_5/\nu_4$; and the preferred dimension $K^*$. The latter is defined as the value of $k$ that maximizes the consecutive quotient-eigenvalue ratio $\nu_{k+1}/\nu_k$ over $k = 2, \dots, 20$. Here, $\nu_4$ and $\nu_5$ are the fourth and fifth eigenvalues of the 42-node quotient Laplacian. The quantity $K^*$ therefore records the number of modes lying below the largest detected separation in this prescribed low-eigenvalue range.

The regional level is highly tolerant of perturbations that preserve efficient mixing within each region (Fig. 5a,b). Lognormal internal-conductance disorder with $\sigma = 0.4$ reduces $g_{42}$ only from 8.08 to 7.01, while $\overline{\chi}$ remains 0.9970. Complete degree-preserving rewiring of the internal edges increases $g_{42}$ to 10.77 and raises $\overline{\chi}$ to 0.9991. The precise microscopic geometry of the internal connections is therefore not essential; what matters is that each region remains strongly and efficiently mixed.

Perturbations that damage internal mixing have a markedly different effect (Fig. 5a,b). Deleting 30% of the internal edges lowers $g_{42}$ to 2.26, while reducing the internal conductance of a single region to 10% of its original value lowers the gap to 1.69. A bottleneck confined to one region can therefore limit the separation of the entire 42-mode band. The regional description fails not because microscopic connectivity has changed in itself, but because the change prevents one or more regions from approaching a common internal voltage.

The global four-mode level responds selectively to perturbations of the interregional network (Fig. 5c,d). Changes confined within regions leave $g_4$ unchanged because they do not modify the quotient boundary graph. Boundary shortcuts have the opposite effect: with 40% shortcut replacement, $g_{42}$ remains relatively large at 7.28, whereas $g_4$ falls from 2.53 to 1.14. At the same time, the preferred global dimension becomes less reproducible across realizations: the fraction for which $K^* = 4$ decreases from 1.00 in the unperturbed network to 0.75 at the strongest shortcut level (Fig. 5d and Supporting Information S5).

These results establish a scale-selective form of robustness. The 42-variable regional level is controlled primarily by mixing within regions, whereas the four-mode global level is controlled by connectivity between regions and by the approximate spherical organization of the quotient network. A collective level can therefore remain insensitive to substantial perturbations below its scale while being strongly altered by changes acting directly at that scale.

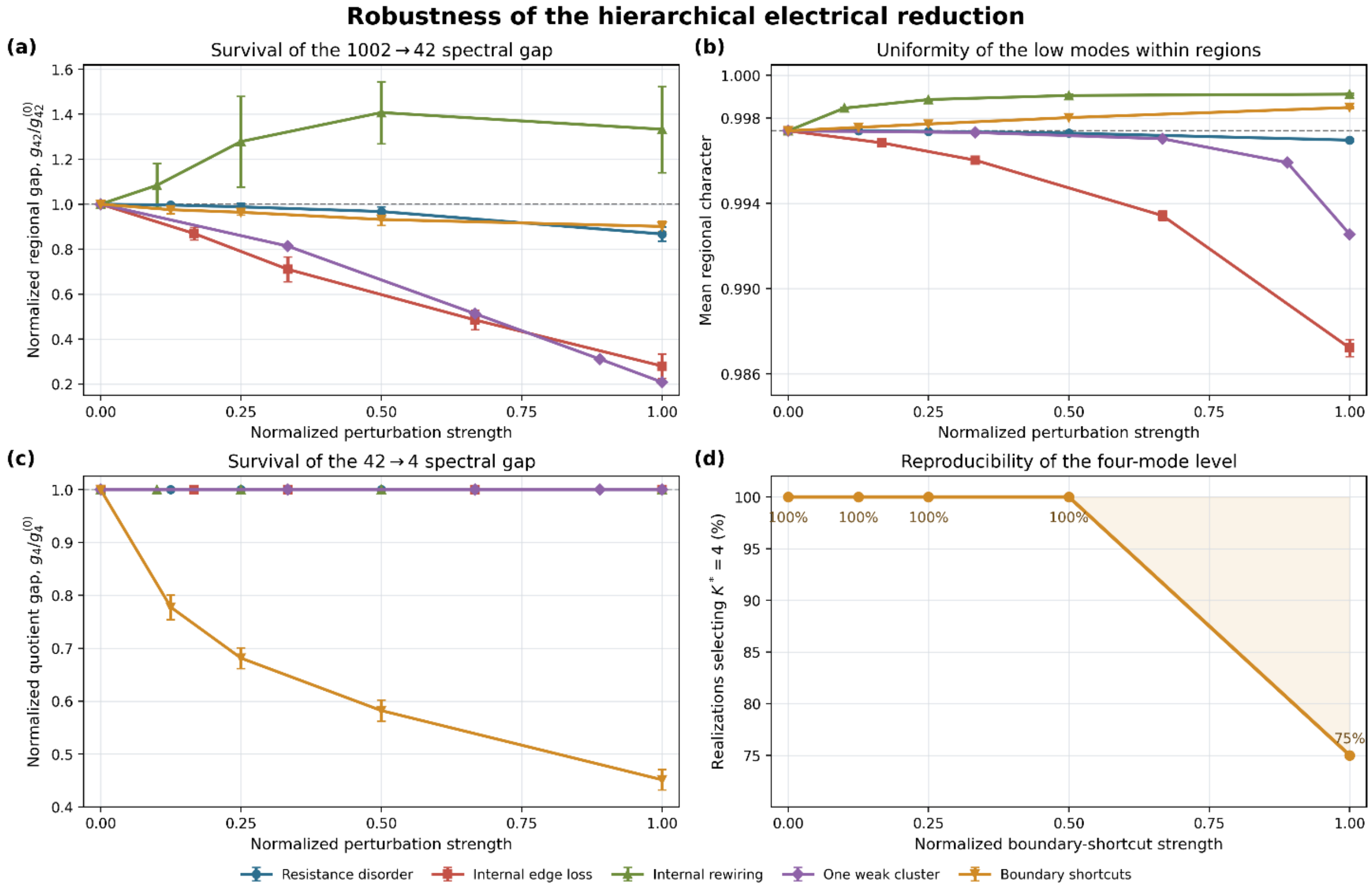

**Figure 5. Robustness of the hierarchical electrical reduction.** (a) Evolution of the normalized regional spectral gap, $g_{42}/g_{42}^{(0)}$, under five classes of perturbation. Each perturbation strength is expressed relative to the largest value tested for that family. Internal-conductance disorder and degree-preserving rewiring preserve, or even strengthen, the regional separation, whereas internal-edge deletion and weakening of a single region progressively close the 42-mode gap. (b) Evolution of the mean regional character, $\overline{\chi}$, obtained by averaging $\chi_m$ over the 41 nonuniform modes $m = 2, \dots, 42$. Here, $\chi_m$ is the fraction of eigenvector variance reproduced by the 42 regional averages; the dashed line indicates the unperturbed value, $\overline{\chi} = 0.9974$. (c) Evolution of the normalized quotient gap, $g_4/g_4^{(0)}$. Only boundary shortcuts substantially weaken the secondary four-mode separation. (d) Percentage of realizations selecting $K^* = 4$ under increasing boundary-shortcut replacement. Agreement decreases from 100% to 75% at the strongest perturbation, indicating that the preferred global dimension becomes less reproducible. Error bars indicate variation across realizations. The superscript $(0)$ denotes the corresponding value in the unperturbed network.

# 4. Constructing effective electrical circuits

## 4.1 One electrical value per region

Let $\mathbf{v} = (v_1, \dots, v_N)^{\mathrm{T}}$ denote the vector of microscopic node voltages, and let $\mathbf{V} = (V_1, \dots, V_M)^{\mathrm{T}}$ contain one collective voltage for each of the $M$ regions.

The $N \times M$ membership matrix $P$ specifies which node belongs to which region:

$$P_{ia} = \begin{cases} 1, & \text{if node } i \text{ belongs to region } a, \\ 0, & \text{otherwise.} \end{cases}$$

The matrix $P$ maps the regional voltages onto the microscopic nodes. Thus, when all nodes within a region have approximately the same voltage,

$$\mathbf{v} \approx P\mathbf{V},$$

so that $v_i \approx V_a$ whenever node $i$ belongs to region $a$. Conversely, multiplication by $P^{\mathrm{T}}$ sums the microscopic Kirchhoff equations over all nodes belonging to the same region. Writing $\mathbf{C}$ and $\mathbf{L}^{-1}$ for the microscopic capacitance and inverse-inductance matrices, $\mathbf{G} = \mathrm{G_0}\, \mathrm{I_N} + \mathrm{G_c}\mathcal{L}$ for the complete nodal conductance matrix, and $I^{ext}(t)$ for the applied-current vector, the projected circuit matrices and input are

$$C_{red} = P^T\mathbf{C}P, \qquad G_{red} = P^T\mathbf{G}P, \qquad L_{red}^{-1} = P^T\mathbf{L}^{-1}P, \qquad I_{red}^{\mathrm{ext}} = P^T I^{ext}(t). \qquad (9)$$

For identical microscopic resonators, let $n_a = \sum_i P_{ia}$ denote the number of nodes in region $a$. The $n_a$ capacitors and leakage conductances contribute in parallel to the regional node, while the $n_a$ identical inductors also form a parallel combination. The effective shunt elements of region $a$ are therefore

$$C_a = n_a C, \quad G_{0a} = n_a G_0, \quad L_a = \frac{L}{n_a}. \qquad (10)$$

Equation (10) defines the shunt elements associated with each regional voltage. The leakage contribution to $P^{\mathrm{T}}\mathbf{G}P$ gives the regional shunt conductance $G_{0a} = n_a G_0$. When Kirchhoff's equations are summed within a region, currents through resistors whose two terminals belong to that same region cancel pairwise. Only currents through edges crossing a regional boundary remain. All microscopic conductances connecting regions $a$ and $b$ act in parallel, giving

$$G_{ab}^{eff} = \sum_{i \in a, j \in b} G_{ij}\,, \quad R_{ab}^{eff} = \frac{1}{G_{ab}^{eff}}. \qquad (11)$$

Here, $G_{ab}^{\mathrm{eff}}$ is the total conductance connecting regions $a$ and $b$, and $R_{ab}^{\mathrm{eff}}$ is the corresponding effective resistance. Thus, for $a \neq b$, $(G_{red})_{ab} = -G_{ab}^{eff}$, while the diagonal element is $(G_{red})_{aa} = n_a G_0 + \sum_{b \neq a} G_{ab}^{eff}$.

Equations (9)–(11) therefore specify every element of the reduced circuit directly from the microscopic component list. No dynamical fitting is involved, and the resulting 42-node network can be physically constructed. Its predictions are compared with the regional averages of the full network, $\overline{v}_a = n_a^{-1} \sum_{i \in a} v_i$.

## 4.2 Validation and the origin of the residual error

For uniform stimulation of the northern region, the 42-variable circuit gives a global root-mean-square regional error of 1.42%. For a point pulse at the north pole and for two separated point pulses, the errors are 4.62% and 4.34%, respectively. The point-pulse error falls below 1% after $28.7\sqrt{LC}$ (Supporting Information S6 and Fig. S4). These results confirm that the quotient circuit captures collective dynamics particularly well when the input is itself coarse-grained.

The most stringent observable is the extremely small voltage reaching the opposite pole. After a northern point pulse, the full network gives a south-region peak of $2.30 \times 10^{-8}$, whereas the 42-variable circuit gives $1.10 \times 10^{-7}$ (Fig. S4a). This large relative discrepancy occurs despite the modest global error. It is consistent with the sensitivity of a weak remote response to the small spectral differences shown in Fig. S4d, although the present calculation does not isolate the individual modal cancellations responsible for the discrepancy. Thus, the spectral gap identifies a useful collective subspace but does not ensure uniform relative accuracy for observables whose amplitude is nearly zero. A complementary frequency-domain test in the 252-node network shows the same limitation for a weak transfer between distant regions (Supporting Information S13 and Fig. S5).

### 4.3 Systematic enrichment by internal polarization modes

The approximation can be improved without returning to all microscopic voltages. For each isolated region, we calculate the eigenvectors of its internal Laplacian. The first eigenvector is uniform; the second is the slowest zero-mean internal deformation. Retaining both gives

$$v_i(t) \approx V_a(t) + A_a(t)\psi_{a,1}(i), \qquad i \in a. \tag{12}$$

$V_a$ is the common regional voltage and $A_a$ is an internal polarization coordinate. Projection of the full circuit matrices onto these 84 orthonormal patterns generates a passive linear reduced system. The global RMS errors fall to 3.19%, 0.91% and 2.69% for the north-point, north-region and two-point stimuli. The point-response settling time below 1% decreases from $28.7\sqrt{LC}$ to $19.35\sqrt{LC}$. The south-pole peak improves from $1.10\times10^{-7}$ to $3.72\times10^{-8}$, close to the full value $2.30\times10^{-8}$.

This calculation changes the status of the reduction. It is not only a heuristic replacement of a cluster by its mean. It is the first member of a systematic hierarchy in which additional internal modes recover directional propagation and boundary polarization. The appropriate reduced dimension can therefore be selected by the required observable and accuracy.

Figure 6 summarizes this comparison. The upper panels show how the 42-variable quotient circuit reproduces the collective north-to-south response and how the inclusion of one internal polarization coordinate per region improves the weak remote signal. The lower panels quantify the corresponding reduction in regional error for the three stimulation geometries. Together, these results show that the spectral hierarchy provides not only a qualitative organization of the dynamics, but also a systematic route for improving the accuracy of the reduced electrical description.

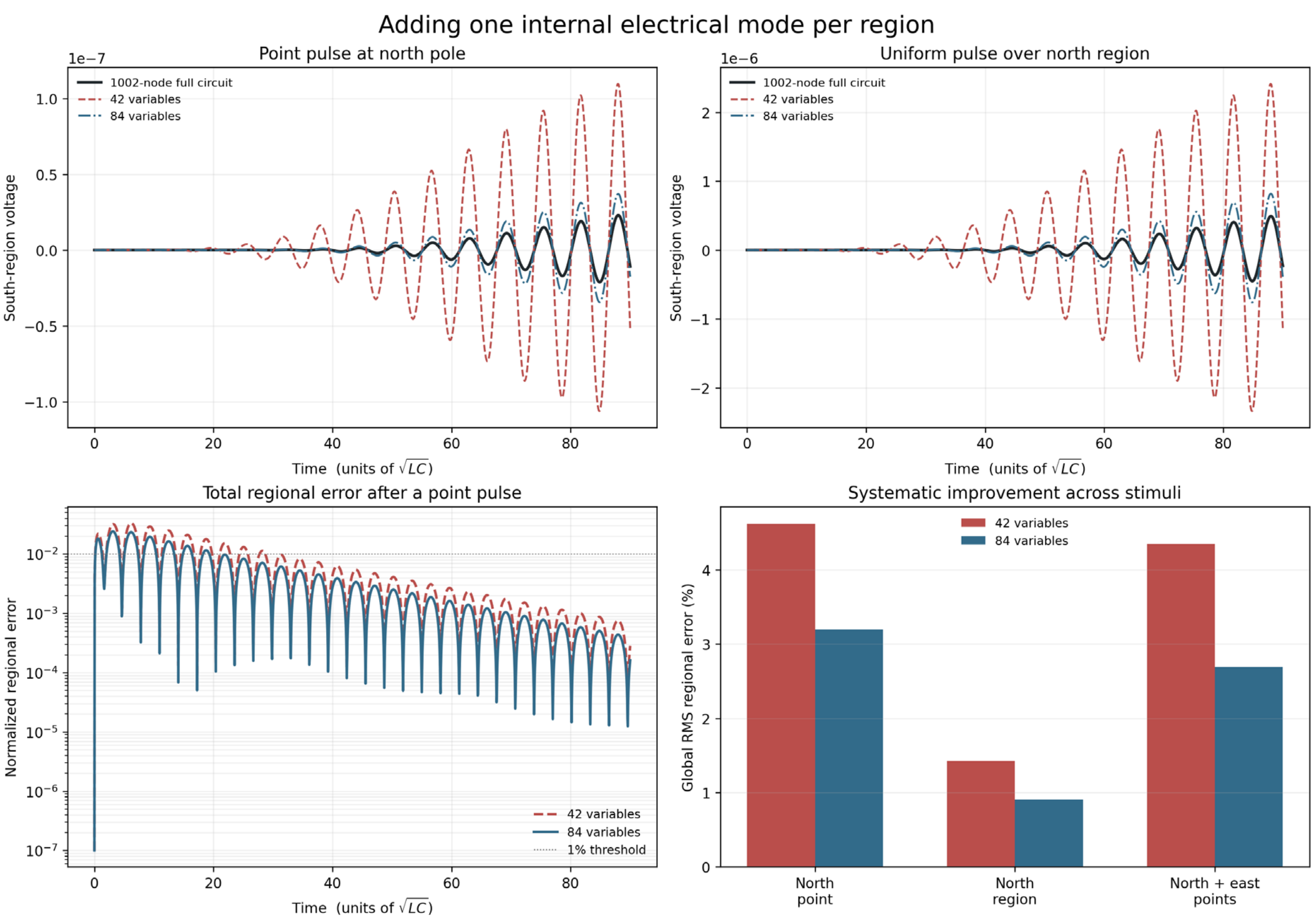

**Figure 6.** Validation and enrichment of the reduced 1002-node circuit. The upper panels compare the difficult north-to-south response of the complete network, the 42-variable regional circuit and the 84-variable model containing one internal polarization per region. The lower panels show the reduction of total regional error and the systematic improvement for three stimulation geometries.

# 5. Generality across network size

## 5.1 Natural levels in the 252-node sphere

Returning to the smaller system permits a more complete progression through the internal spectrum. The dominant spectral gap occurs after mode 12, with $\frac{\mu_{13}}{\mu_{12}} = 9.42$. A secondary global gap appears after mode 4 with $\frac{\mu_5}{\mu_4} = 2.01$. In addition, a weaker candidate gap occurs after mode 36 with $\frac{\mu_{37}}{\mu_{36}} = 1.37$. These three spectral boundaries are identified in Fig. 7a, where the full spectrum is divided into the 4-, 12-, and 36-dimensional candidate subspaces. Inspection of the local eigenvectors identifies the 36-dimensional band as three coordinates per region: one common voltage and two internal polarization patterns. The corresponding eigenvector structures are illustrated for representative modes in Supporting Information Section S3 (Fig. S2). The resulting hierarchy is

$$252 \text{ microscopic voltages} \rightarrow 36 \text{ regional coordinates} \rightarrow 12 \text{ regional voltages} \rightarrow 4 \text{ global modes.} \qquad (13)$$

Figure 7c illustrates this hierarchy dynamically: after the rapid decay of intraregional differences, the individual node voltages approach their regional averages, while the remaining slower evolution is governed by interregional and global modes. The response calculations confirm the hierarchy while also showing that accuracy improves continuously beyond the clearest gaps. For a northern point pulse, the global RMS error is 6.14% with 12 variables, 3.96% with 24, 2.75% with 36, 1.39% with 48, 0.76% with 60, 1.18% with 72, 0.36% with 84 and 0.24% with 96 (the 72-variable point is the mild, discussed non-monotonicity of Fig. 7). The complete error sequence is plotted in Fig. 7b, while Fig. 7c shows how the progressive inclusion of internal regional modes recovers the remote north-to-south response. The definition of the RMS error and the stimulation protocol are given in Supporting Information Section S6, and the corresponding numerical data for the 252-node network are provided in Supporting Information Section S8. The exact south-region peak is $3.47\times10^{-5}$; the 12-variable model predicts $5.01\times10^{-5}$, the 36-variable model $3.24\times10^{-5}$, and the 84-variable model $3.51\times10^{-5}$.

The error is not strictly monotonic when equal numbers of local modes are added to every region: the 72-variable point-stimulation model is slightly less accurate than the 60-variable model. This small reversal can be seen directly in the local increase of the error curve in Fig. 7b (and in the annotated bar chart of Fig. 7d); it is therefore a feature of this particular equal-allocation sequence rather than a breakdown of convergence of the Galerkin approximation. This does not violate Galerkin convergence in the energy norm; it shows that a particular regional observable over a finite time interval need not improve monotonically under a locally ranked basis. An optimized implementation should select internal modes globally using eigenvalue, input overlap and output observability rather than imposing the same rank in every region. The enriched regional reductions used for this comparison, including the construction of the 24–96-variable bases, are described in Supporting Information Sections S7 and S8.

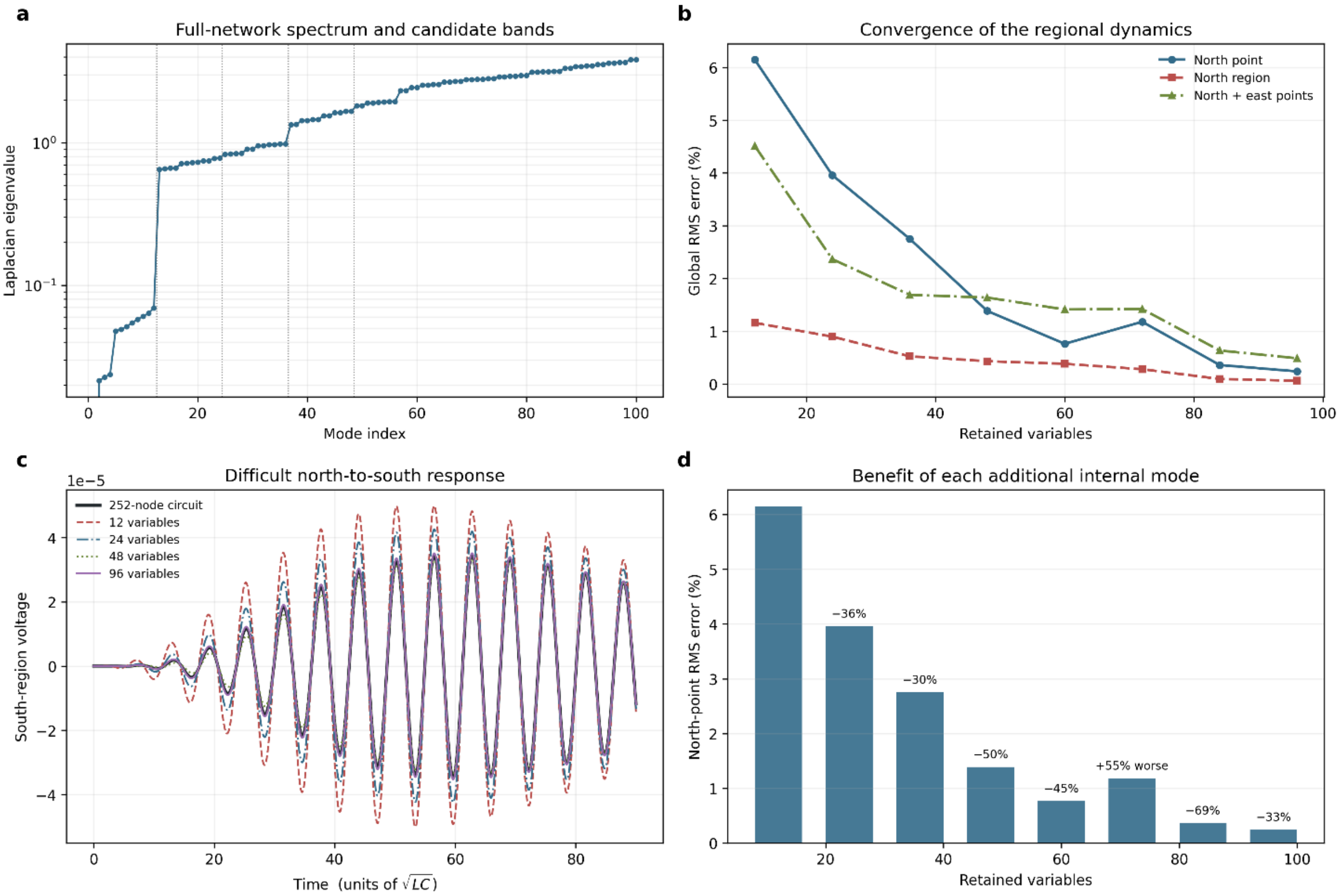

**Figure 7. Multilevel reduction of the 252-node network.** (a) The ordered Laplacian spectrum exhibits candidate separations after modes 4, 12 and 36. (b) Global root-mean-square regional errors for three excitation patterns as the number of retained variables increases from 12 to 96. (c) Progressive recovery of the weak north-to-south response by successive reduced models. (d) North-point error and the relative improvement produced by each additional set of regional internal modes. The mild nonmonotonicity at 72 variables shows that assigning the same number of modes to every region does not necessarily improve a specific observable monotonically, motivating response-dependent mode selection.

## 5.2 General conclusion from both systems

The two networks have different microscopic sizes and different intermediate spectra, but they obey the same organizing principle. Strong intraregional conductance produces a low subspace whose eigenvectors are almost uniform within regions. The quotient of those regions inherits a still lower set of global patterns controlled by the large-scale topology. Additional spectral bands represent internal regional polarization. The numerical values 4, 12, 36 and 42 are system-specific; the correspondence between spectral bands and effective electrical variables is general. This comparison is supported by the reduction sequences obtained for the two network sizes: the 1002-node system develops the $1002 \rightarrow 42 \rightarrow 4$ hierarchy described in Section 3, whereas the 252-node system exhibits the more finely resolved $252 \rightarrow 36 \rightarrow 12 \rightarrow 4$ hierarchy shown in Figs. 7 and 8.

The robustness experiments clarify the meaning of this generality. A complementary robustness dimension — tolerance to component (capacitance and inductance) dispersion — is reported for the 252-node network in Supporting Information Section S10, while the non-embedded random-regular modular network examined in Supporting Information Section S11 shows that the regional mechanism does not depend on the spherical embedding. It is not invariance to arbitrary rewiring. Rather, it is conditional and multiscale. A level survives disorder below it when the property responsible for separation—such as rapid internal mixing—is retained. It is altered by perturbations that weaken that mixing or modify connectivity at the same organizational level. Specifically, the Supporting Information results show whether the relevant eigenvalue gaps, regional eigenvector localization, and reduced-response accuracy survive each class of perturbation. This provides an operational criterion for when coarse-graining is physically justified and when it fails.

### 5.3 From spectral bands to electrical descriptions

The central construction is summarized in Fig. 8: panel (a) enlarges the low-mode region of the spectrum and identifies the boundaries after modes 4, 12, and 36, while panel (b) shows the complete spectrum of the 252-node circuit. The complete spectrum of the 252-node circuit is shown together with an expanded view of its low-mode region, where the relevant spectral boundaries can be distinguished. The strongest separation occurs after mode 12, with $\mu_{13}/\mu_{12} = 9.42$, defining a subspace whose eigenvectors are nearly uniform within each of the twelve regions. Two additional, weaker separations occur after modes 4 and 36. These boundaries identify successive levels of spatial resolution. The first four modes retain the uniform mode and three approximately directional global patterns. These four spatial patterns —the first the uniform collective voltage, the following three the lowest-order directional variations across the spherical network— are analysed in Supporting Information Section S3 (Fig. S2). The first twelve modes assign one collective voltage to each region. Extending the basis to 36 modes introduces two internal polarization patterns per region, whereas retaining all 252 modes restores the complete microscopic voltage distribution. The transition from the four global patterns to the twelve regional common-voltage coordinates, and subsequently to the 36-coordinate basis, is indicated by the spectral bands in Fig. 8a. Representative spatial eigenvectors from each band are analysed in Supporting Information Section S3.

These spectral levels have different physical meanings. The 12-variable description is a directly constructible RLC quotient circuit obtained by summing the microscopic Kirchhoff equations within each region. Its capacitances, inductances and interregional conductances follow directly from the original component values. The aggregation of the microscopic Kirchhoff equations and the resulting quotient matrices are derived explicitly in Supporting Information Sections S4 and S6. By contrast, the 36- to 96-variable descriptions are enriched Galerkin models. They retain additional internal eigenvectors and therefore resolve progressively finer intraregional voltage structure. Their construction and numerical convergence are documented in Supporting Information Sections S7 and S8 and quantified by the response-error curve in Fig. 7b. Their projected matrices are symmetric and positive semidefinite, so the reduced dynamics remains passive. However, the presence of signed off-diagonal elements means that these modal systems are not generally equivalent to simple resistor graphs in which every coordinate is represented by an ordinary circuit node connected only through positive resistances (Supporting Information S12). They should instead be interpreted as passive reduced multiport models, suitable for numerical implementation or more general network synthesis. This distinction is essential: passivity of the projected dynamics does not imply that every reduced matrix can be realized as a simple graph containing only pairwise positive resistors.

The separation after mode 36 is modest, $\mu_{37}/\mu_{36} = 1.37$, and is therefore identified as a candidate polarization level rather than a sharply defined or universal dimension. The spectral boundaries should consequently not be interpreted as prescribing a unique reduced model. They identify physically meaningful subspaces at different resolutions. The operational dimension must then be selected according to the spatial structure of the excitation, the measured observable and the required accuracy. This dependence is demonstrated in Fig. 7c,d: the number of modes required to reproduce a local or remote response is not fixed solely by the spectral gaps, but also by the overlap of the retained eigenvectors with the applied excitation and the measured output. Frequency-domain confirmation of this conclusion is provided in Supporting Information Section S13.

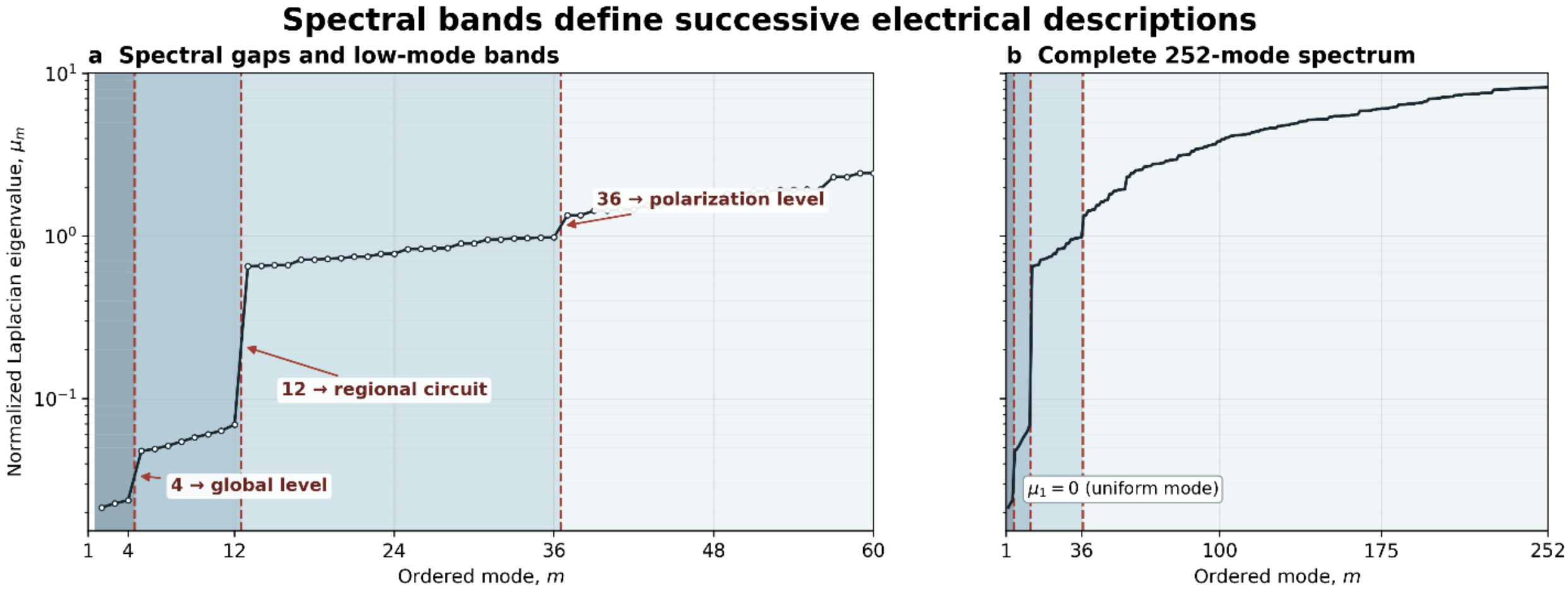

**Figure 8. Spectral bands and successive electrical descriptions of the 252-node network.** The figure summarizes the hierarchy obtained for the spherical network of 252 microscopic circuit nodes partitioned into 12 regions. (a) Expanded view of the first 60 eigenvalues, showing spectral gaps after modes 4, 12 and 36. These boundaries identify, respectively, a four-mode global subspace, the 12 regional voltages defining the directly constructible quotient circuit, and a candidate 36-coordinate level that includes two internal polarization coordinates per region. (b) Complete ordered spectrum of the same 252-node circuit, showing how these reduced descriptions occupy the lower part of the full microscopic spectrum. Reading the spectrum from low to high mode index corresponds to increasing spatial resolution. Only the one-voltage-per-region description is automatically a simple resistor-coupled RLC graph; the enriched modal descriptions are passive Galerkin systems whose physical realization may require a general multiport network.

## 6. Discussion

Graph coarse-graining is usually formulated as an approximation problem: construct a smaller graph that preserves selected cuts, spectra or dynamical responses, including for modular and small-world topologies [17–22,33–36]. The present work adds a complementary physical interpretation. The projection matrices are not only algebraic maps; they define voltages that can be measured, effective capacitances and inductances that can be assembled, and resistor values obtained by ordinary parallel combination. This makes it possible to test spectral reduction as a circuit, rather than only compare matrices.

It is worth contrasting this explicitly with Kron reduction [34], which eliminates internal nodes exactly via the Schur complement but preserves every retained node individually, and with spectral graph-coarsening algorithms that guarantee bounded spectral or cut distortion for a chosen reduction ratio [20–22]. The present approach differs in aim rather than in rigour: instead of eliminating nodes exactly or bounding a generic distortion metric, it seeks the smallest number of physically meaningful aggregate variables – one per strongly mixed region – and accepts a controlled, quantified approximation error in exchange for a circuit that is an order of magnitude smaller and directly buildable. A systematic comparison of the achievable error at matched reduction ratios against algorithmic coarsening is a natural next step.

A practical distinction is important. The one-voltage-per-region reduction maps directly onto component values: capacitors and leakage conductances add, parallel inductors give $L_a = L/n_a$, and boundary conductances add between regions. No transfer-function fitting is involved, and this quotient can be assembled from ordinary components. Enriched modal reductions are also derived without fitting and remain passive, but they are reduced multiport systems rather than automatically simple resistor graphs. The purpose is diagnostic: the spectrum identifies candidate collective subspaces, the eigenvectors identify their voltage patterns, and response validation selects the smallest description adequate for a specified measurement.

The results also sharpen the role of eigenvalues. A small eigenvalue does not mean that a mode has small amplitude; amplitude depends on input overlap and observation. It means that the corresponding voltage pattern produces little dissipation in the coupling network. Similarly, a spectral gap is not by itself a universal timescale separation because the shunt dynamics, forcing frequency and damping also enter Eq. 5. It is evidence for a separated family of spatial conductance penalties. The time-domain validation is therefore indispensable.

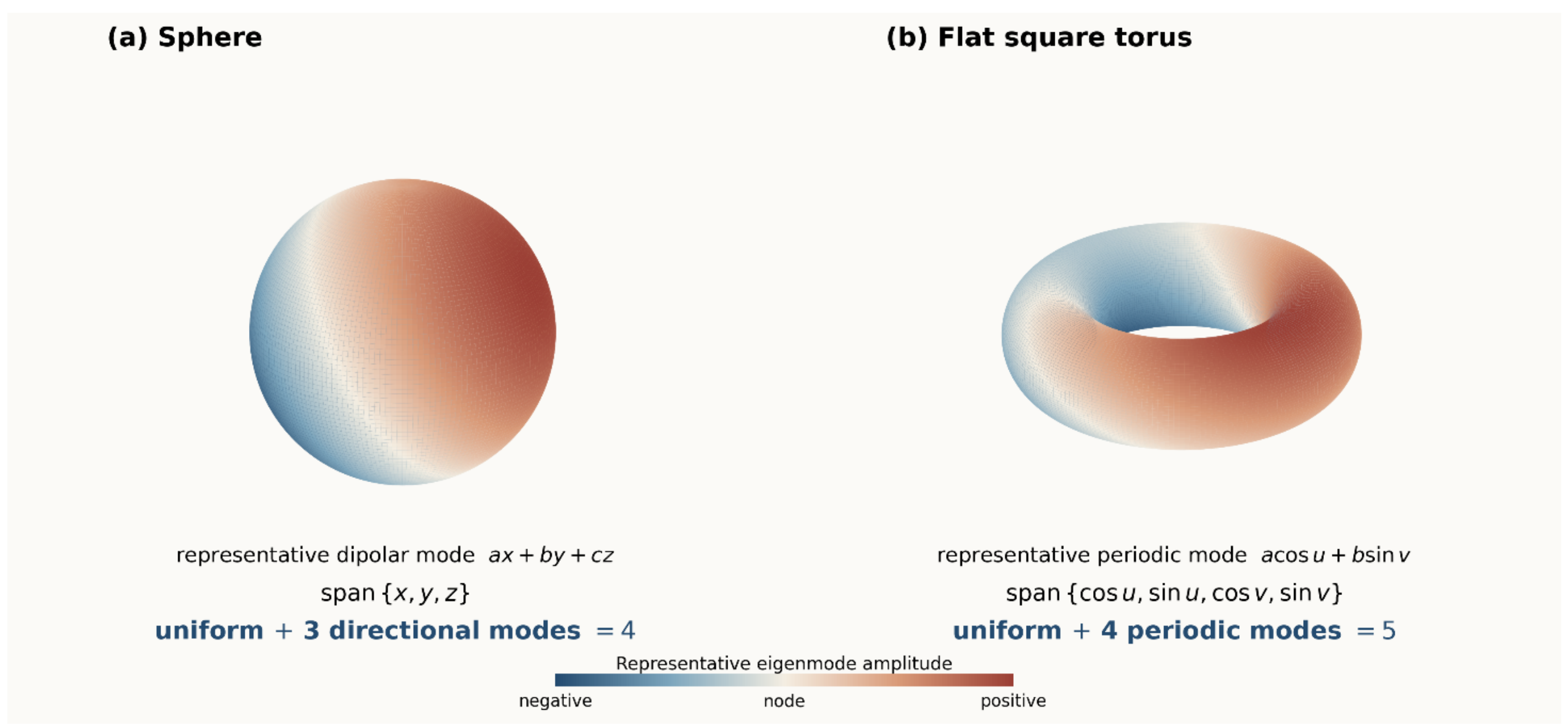

**Figure 9. Geometry selects the dimension of the lowest Laplacian eigenspace. (a)** A representative dipolar mode on the sphere, expressed as a linear combination ax + by + czax + by + cz. The diffuse colour scale represents the signed eigenmode amplitude, with blue and red denoting opposite signs and the light region indicating the nodal transition. The three coordinate functions xx, yy, and zz span the first nonzero spherical-harmonic eigenspace. Together with the uniform zero mode, they produce a four-dimensional lowest global subspace. **(b)** Representative periodic mode on a flat square torus, expressed here as a linear combination of $\cos u$ and $\sin v$. Its complete lowest nonzero eigenspace is spanned by $\cos u$ , $\sin u$, $\cos v$ and $\sin v$. Together with the uniform mode, this produces a five-dimensional lowest global subspace. The doughnut provides a schematic visualization of the toroidal topology; the stated fourfold degeneracy corresponds specifically to the ideal symmetric flat square torus. The comparison demonstrates that the four-mode level observed in the spherical networks is selected by geometry and symmetry rather than being a generic property of modular graphs.

The four lowest modes deserve particular attention. A connected network always has one uniform zero mode. The additional three modes arise here because the interregional graphs approximate a homogeneous two-dimensional spherical surface. As illustrated in Fig. 9a, the first nonzero eigenspace of the Laplace–Beltrami operator on a sphere is the three-dimensional ($\ell = 1$) spherical-harmonic subspace, which can be represented by the coordinate functions (x), (y), and (z). The graded colour pattern in Fig. 9a shows one representative dipolar mode ($ax + by + cz$) with opposite signs separated by a nodal region. Because the three modes are nearly degenerate, the numerical graph eigenvectors need not coincide individually with (x), (y), and (z), but may appear as rotated linear combinations spanning approximately the same directional subspace. Together with the uniform mode, this spherical geometry therefore produces a four-dimensional lowest global level.

This dimensionality is geometry-dependent rather than a generic consequence of modular organization. For comparison, Fig. 9b illustrates an ideal flat square torus, whose two periodic coordinates, (u) and (v), each contribute a sine–cosine pair. Its lowest nonzero eigenspace is therefore spanned by $\cos u$ , $\sin u$, $\cos v$ and $\sin v$, giving five lowest global modes when the uniform mode is included. The torus in Fig. 9b is a schematic representation of the periodic topology; the fourfold degeneracy applies specifically to the symmetric flat square torus and can be split by geometric distortion or anisotropy. More generally, surface topology, boundary conditions, symmetry and anisotropy determine the dimensions and degeneracies of the lowest global eigenspaces.

The recurrence of four modes in both spherical network sizes is thus a geometric result, whereas the separation between regional and internal modes is produced by modular conductance contrast. As a control, twenty non-embedded modular networks formed from independent 10-regular regions retain a sharp regional band but show no stable low-dimensional quotient gap (Supporting Information Section S11). This control supports the distinction between the general modular mechanism that generates regional collective variables and the geometry-specific mechanism that selects the four-mode global structure.

The analogy with brain dynamics must remain limited. Large neural systems also show hierarchical modularity, low-dimensional population activity and connectivity-constrained modes [24–32]. However, neuronal ensembles are nonlinear, active, heterogeneous, delayed and adaptive. Our results neither model neuronal synchronization nor identify neural functional assemblies. They provide a transparent physical example of how hierarchical connectivity alone can produce nested collective variables. This may be useful as a controllable reference system for reasoning about more complex networks without conflating analogy with biological identity.

Several extensions follow naturally. The component-tolerance calculation in Supporting Information S10 shows that moderate dispersion in $C$ and $L$ degrades accuracy gradually, while also breaking exact diagonalization by the Laplacian eigenvectors. Delays, capacitive or inductive links would make the coupling operator frequency dependent. Nonlinear or active resonators would permit self-oscillation and genuine synchronization, but would also require local stability and amplitude dynamics beyond Laplacian diagonalization. Experimentally, the reduced directly aggregated networks could be assembled on printed circuit boards, with programmable resistor arrays used to test topology changes in real time.

## 7. Conclusions

Resistor-coupled LC networks make the graph Laplacian directly physical. Its eigenvectors are voltage patterns and its eigenvalues are the coupling conductances that damp those patterns. Spectral gaps separate collective electrical levels: one voltage per strongly mixed region, internal regional polarizations, and global modes imposed by large-scale geometry. The 1002-node and 252-node spheres demonstrate different hierarchies governed by the same principle. Directly aggregated circuits reproduce collective evolution, and local modal enrichment progressively recovers information lost by averaging. The conclusion is compact: eigenvalues identify candidate dimensions, eigenvectors show what their variables mean, and response validation determines whether a reduction is adequate for the observable of interest. The regional mechanism does not require spherical geometry, but the four-mode global level does. The directly aggregated quotient circuits are realizable from ordinary components; enriched modal models require a more general passive-network interpretation.

## Acknowledgments

The author acknowledges support from the European Research Council (ERC) via Horizon Europe Advanced Grant, grant agreement nº 101097688 ("PeroSpiker"). The numerical exploration, organization of the analysis, preparation of figures and drafting of the manuscript were assisted by OpenAI ChatGPT/Codex, and the revision of initial versions— including an independent verification of the derivations and numerical results and an initial non-embedded modular-network control — was assisted by Claude (Anthropic). Both assisted at the request and under the scientific direction of Juan Bisquert, who defined the research questions, selected the physical models, evaluated the interpretations and assumes full responsibility for the scientific content. No generative AI system is listed as an author.

## Data and code availability

The data presented here can be accessed at https://doi.org/10.5281/zenodo.22670423 (Zenodo) under the license CC-BY-4.0 (Creative Commons Attribution 4.0 International).

**Supporting Information**

# Making the Graph Laplacian Physical: Multiscale Coarse-Graining in Electrical Oscillator Networks

## Juan Bisquert

Instituto de Tecnología Química (ITQ), Consejo Superior de Investigaciones Científicas-Universitat Politècnica de València, Valencia, Spain.

Corresponding author email: jbisquer@itq.upv.es

## S1. Network construction

The spherical graphs were generated from an icosahedron. Every triangular face was subdivided with frequency f, duplicate vertices on shared edges were merged, and all coordinates were normalized to unit radius. Frequencies $f = 5$ and $f = 10$ produce $N = 10f^2 + 2 = 252$ and 1002 vertices, respectively. Edges are inherited from the triangular subdivision, giving locally sixfold coordination except at the twelve topological defects required on a sphere.

Regional centres were obtained from lower-frequency geodesic spheres. The 252-node graph used the 12 icosahedral directions; the 1002-node graph used 42 directions from a frequency-2 geodesic sphere. A small rigid rotation was applied to the reference directions to remove exact assignment ties. Each microscopic node was assigned to the centre maximizing the dot product of their unit vectors. This is a spherical Voronoi partition.

Every triangulation edge inside one region was assigned normalized conductance 1. Every boundary edge was assigned conductance $r = \frac{R_s}{R_L}$. Unless otherwise stated, $r = 0.03$. The physical reference values used in the earlier dimensional illustrations were $R_s = 100\ \Omega$, $C = 10\ \mu\text{F}$, $L = 10$ mH and $R_0 = 5\ \text{k}\Omega$; time-domain comparison in the main paper used normalized $C = L = 1$ and $G_0 = 0.08$ to display the modal decay clearly. These two parameter sets serve different purposes and are not meant to correspond numerically: the physical values confirm that the circuit can be built from ordinary components, while $G_0 = 0.08$ is chosen purely for clarity of the modal-decay plots (with $R_0 = 5\ \text{k}\Omega$ and $R_s = 100\ \Omega$ as in the physical set, the implied normalized $G_0$ would be about 0.02, not 0.08).

## S2. Circuit equations and modal solution

For node voltages v, inductor currents j and external currents I, the first-order state equations are

$$C\frac{dv}{dt} = I - G_0 v - G_c \mathcal{L} v - j, \tag{S1}$$

$$L\frac{dj}{dt} = v. \tag{S2}$$

For identical scalar C and L, the eigenvectors of $\mathscr{L}$ diagonalize the full system. Writing $v = \Phi q$ and $j = \Phi s$, with $\Phi^{\mathrm{T}}\Phi = I$ and $\Phi^{\mathrm{T}}\mathcal{L}\Phi = \mathrm{diag}(\mu_m)$, gives independent two-state systems

$$C\frac{dq_m}{dt} = I_m - (G_0 + G_c\mu_m)q_m - s_m, \qquad L\frac{ds_m}{dt} = q_m. \tag{S3}$$

For normalized $C = L = G_c = 1$, zero input after t=0, $q_m(0) = q_m^0$ and $s_m(0) = 0$, the voltage satisfies

$$\frac{d^2q_m}{dt^2} + a_m\frac{dq_m}{dt} + q_m = 0, \qquad a_m = G_0 + \mu_m. \tag{S4}$$

Let $\rho_{m,\pm} = \frac{\left(-a_m \pm \sqrt{a_m^2 - 4}\right)}{2}$. The exact kernel used in the simulations is

$$h_m(t) = \frac{[(-a_m - \rho_{m,-})\exp(\rho_{m,+}t) + (\rho_{m,+} + a_m)\exp(\rho_{m,-}t)]}{(\rho_{m,+} - \rho_{m,-})}, \tag{S5}$$

so that $v(t) = \Phi\ \mathrm{diag}[h_m(t)]\ \Phi^{\mathrm{T}} v(0)$. Complex conjugate roots were evaluated in complex arithmetic and the real part was retained. This modal calculation avoids time-stepping error.

## S3. Spectral diagnostics

The regional character of eigenvector $\varphi_m$ was defined as the fraction of its variance explained by regional means. If $n_a$ is the number of nodes in region a and $\bar{\varphi}_{ma}$ is the mean over that region,

$$\chi_m = \frac{[\sum_a n_a(\bar{\varphi}_{ma} - \bar{\varphi}_m)^2]}{[\sum_i (\varphi_{mi} - \bar{\varphi}_m)^2]}. \tag{S6}$$

$\chi_m = 1$ denotes a perfectly piecewise-constant regional pattern; $\chi_m \approx 0$ denotes a mode whose variation is internal to regions. For K regions, the modular spectral gap was measured as $g_K = \frac{\mu_{K+1}}{\mu_K}$. The secondary global gap was calculated from the generalized eigenvalues of the regional quotient network.

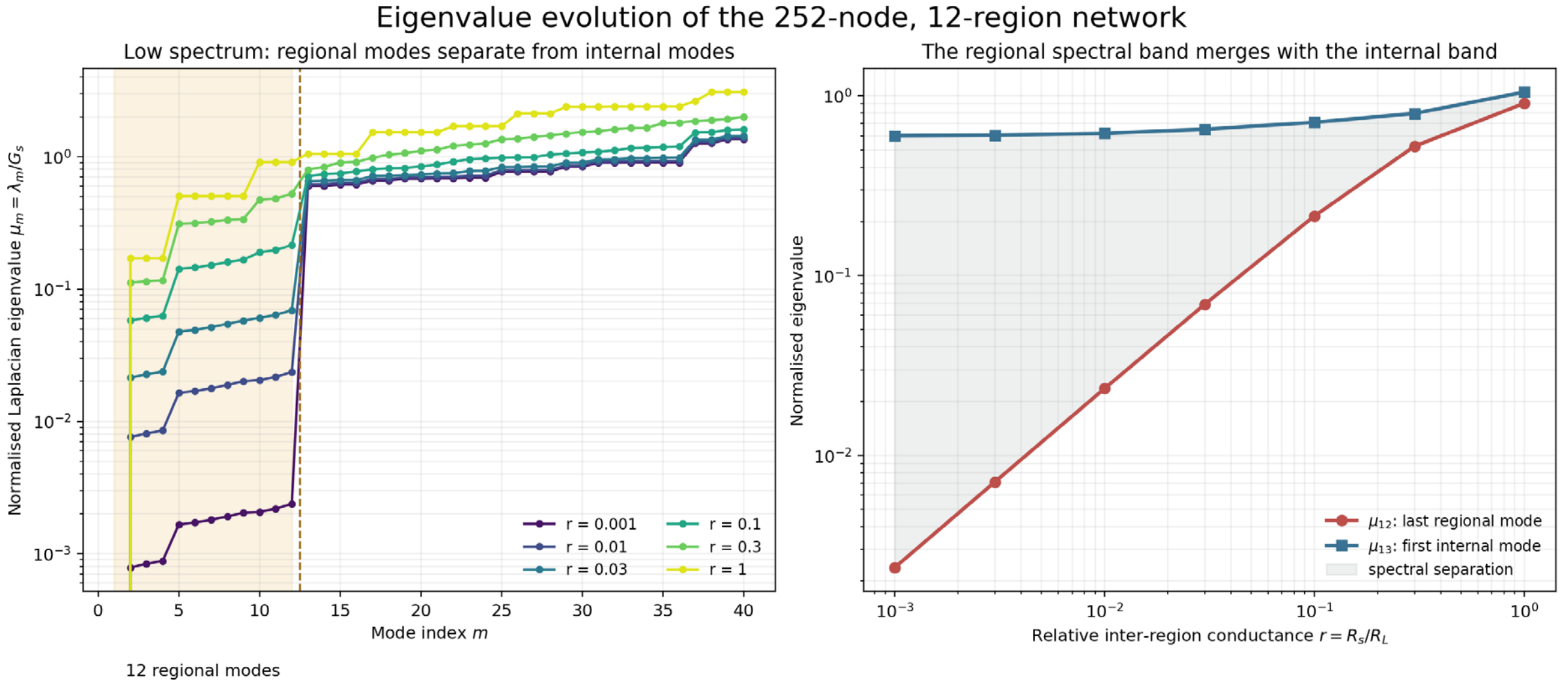

**Figure S1.** Complete low-spectrum analysis of the 252-node network as r is varied. The first 12 eigenvalues form a regional band at weak boundary coupling. The right panel quantifies closure of the μ13/μ12 separation as interregional conductance increases.

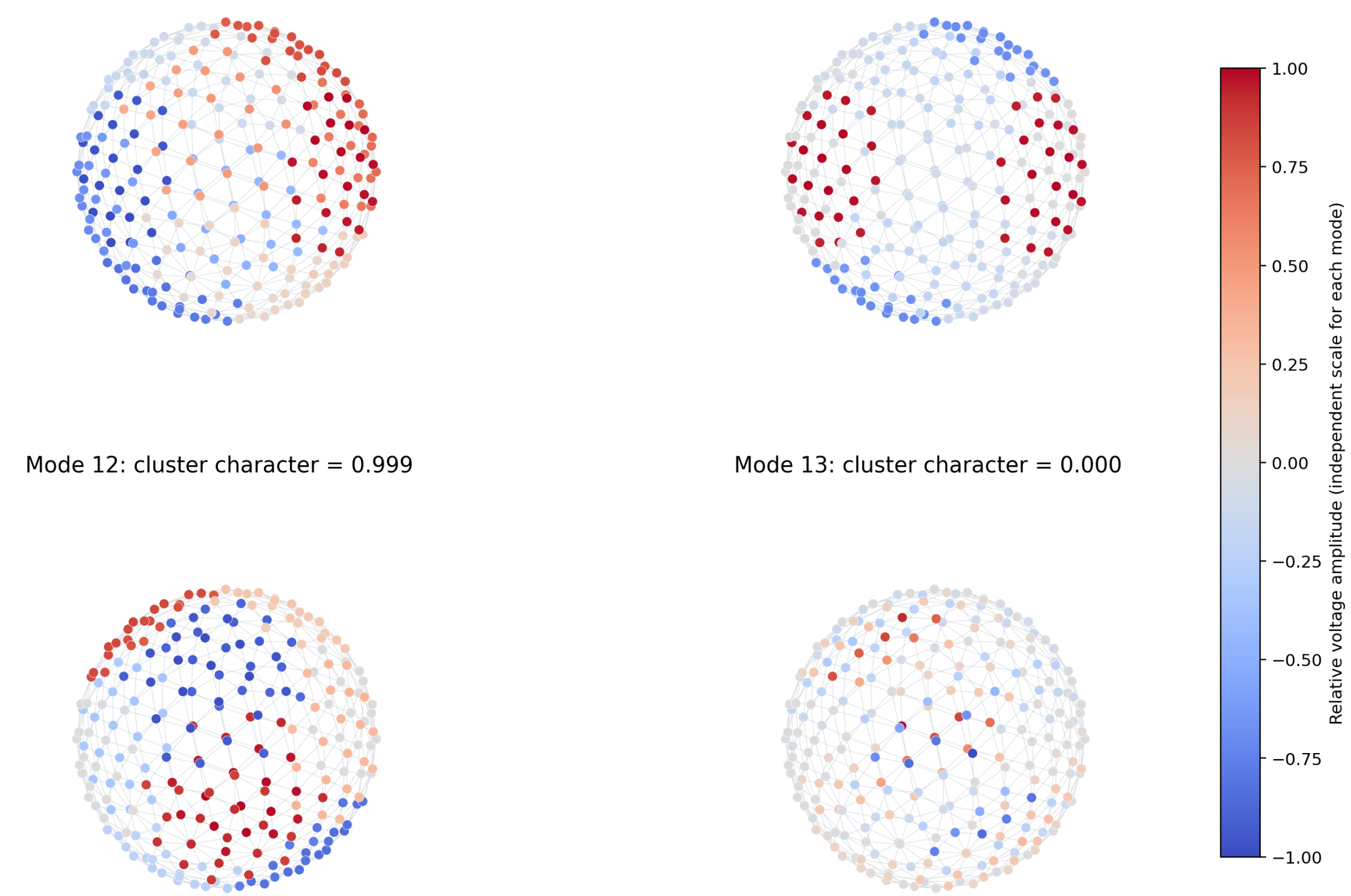

**Figure S2.** Representative eigenvectors of the 252-node network at $r = 0.03$. Modes below the 12-mode gap have high cluster character, while mode 13 is the first to show substantial internal regional variation.

## S4. Quotient network and natural groups

The binary membership matrix P has $P_{ia} = 1$ if microscopic node i belongs to region a. The diagonal size matrix is $M = P^{\mathrm{T}}P = \mathrm{diag}(n_a)$. Direct aggregation of the coupling Laplacian gives $\mathcal{L}_q = P^{\mathrm{T}}\mathcal{L}P$. The regional spectrum is obtained from the generalized eigenproblem

$$\mathcal{L}_q u_a = \nu_a M u_a. \qquad \text{(S7)}$$

The generalized normalization is required because the regional nodes contain unequal numbers of microscopic resonators. The quotient conductance between regions a and b equals the sum of all microscopic boundary conductances between them. The same construction underlies the physical 42-node circuit.

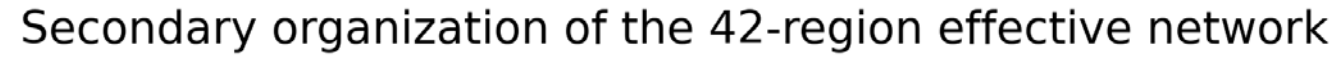

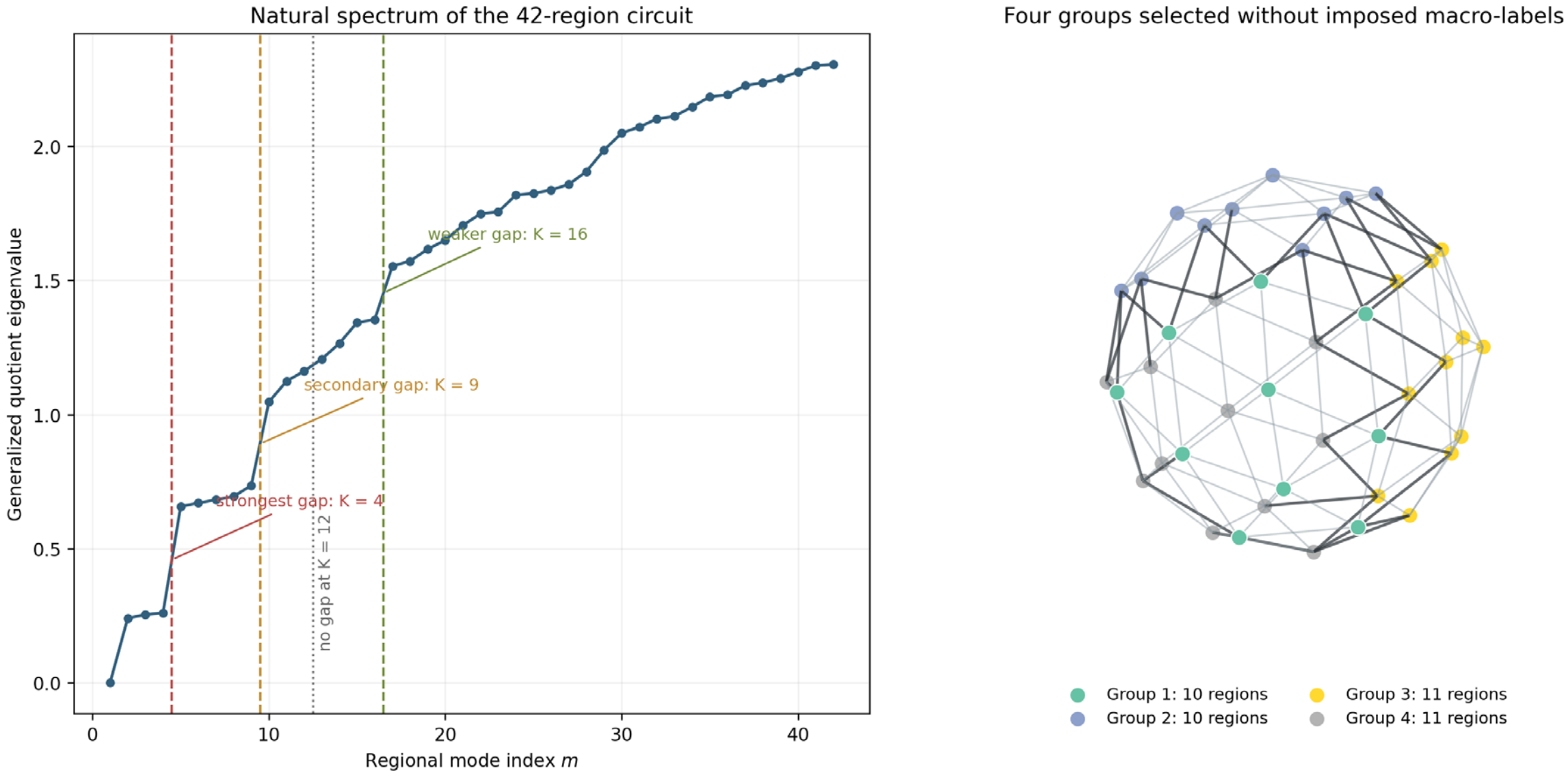

**Figure S3.** Natural organization of the 42-region quotient network. Low regional eigenvectors and their correlations select four global coordinates. The colours identify the macroscopic grouping implied by the low spectral subspace rather than an additional imposed partition.

## S5. Robustness perturbations

Five perturbation families were evaluated at r=0.03. Each nonzero random condition used eight independent realizations. Resistance disorder multiplied internal conductances by lognormal factors $\exp\left(\sigma\xi - \frac{\sigma^2}{2}\right)$, preserving the mean conductance. Internal edge loss removed a prescribed fraction while retaining a random spanning tree in every region. Internal rewiring used double-edge swaps that preserve microscopic degree and was rejected if a region became disconnected. The weak-region test multiplied every internal conductance of the largest region by q. Boundary shortcuts replaced a fraction of boundary edges by links between randomly selected distant regions while maintaining the total boundary-edge count.

The observables were $g_{42} = \frac{\mu_{43}}{\mu_{42}}$, mean $\chi_m$ over modes 2–42, the quotient gap $g_4 = \frac{\nu_5}{\nu_4}$, and the position K* of the strongest low-spectrum quotient gap. The baseline values were $g_{42} = 8.083$, mean $\chi = 0.997400$ and $g_4 = 2.525$. At the strongest tested perturbation, $\sigma = 0.4$ gave $g_{42} = 7.010 \pm 0.258$; 30% internal edge loss gave 2.265±0.438; complete internal rewiring gave 10.774±1.549; and weakening one region to $q = 0.1$ gave 1.686. Forty per cent boundary shortcuts preserved $g_{42} = 7.283 \pm 0.188$ but lowered $g_4$ to 1.140±0.049. Boundary shortcuts also lowered the fraction of realizations that agree on the strongest secondary gap K* (preferred_K_fraction) from 1.00 at baseline to 0.75 at the strongest level tested, while every other perturbation family left this fraction at 1.00; this is the quantity plotted against marker size and as a dashed line in Fig. 5.

## S6. Direct 42-node physical reduction

Under the piecewise-constant approximation $v \approx PV$, projection gives $C_{\mathrm{red}} = P^{\mathrm{T}}CP$, $G_{\mathrm{red}} = P^{\mathrm{T}}GP$ and $I_{\mathrm{red}} = P^{\mathrm{T}}I$. For $n_a$ identical parallel resonators this gives $C_a = n_a C$, $G_{0a} = n_a G_0$ and $L_a = \frac{L}{n_a}$. Internal resistor currents cancel. All microscopic conductances crossing a pair of regions add in parallel.

The full regional average was $\bar{v} = M^{-1}P^{\mathrm{T}}v$. The normalized time-dependent error was

$$\varepsilon(t) = \frac{\left\| M^{\frac{1}{2}}[\bar{v}(t) - V_{red}(t)] \right\|_2}{\left\| M^{\frac{1}{2}}\bar{v}(0) \right\|_2}. \tag{S8}$$

The global RMS relative error was calculated by integrating the squared numerator and denominator over the complete simulated interval. A settling time below threshold η was defined as the first time after which ε(t) remained below η.

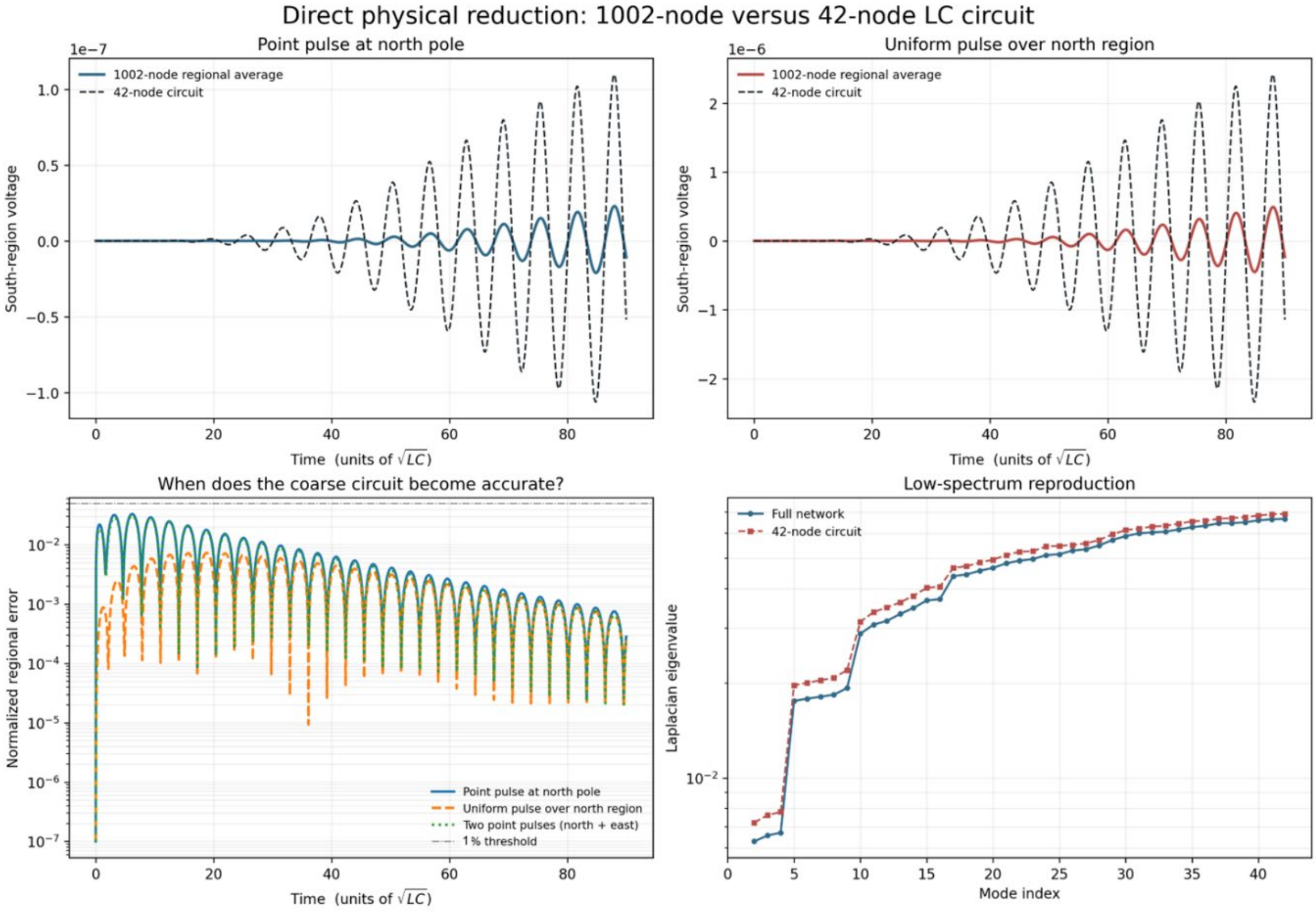

**Figure S4.** Direct validation of the 42-node circuit. The quotient model gives a small total regional error but overestimates the exceptionally weak signal at the opposite pole. The low eigenvalues are reproduced approximately rather than exactly, explaining accumulated phase and damping differences.

## S7. Enriched regional bases

For each isolated region, the eigenvectors of its internal Laplacian were calculated. The uniform vector was fixed as the first normalized basis vector. Subsequent local eigenvectors have zero regional mean and describe increasingly fine internal voltage patterns. A basis retaining q vectors per region contains qM columns Bq and yields the Galerkin matrices

$$C_q = B_q{}^T C B_q, \quad \mathcal{L}_q = B_q{}^T \mathcal{L} B_q, \quad I_q = B_q{}^T I. \tag{S9}$$

Because the local bases have disjoint support and are orthonormal within each region, $B_q{}^{\mathrm{T}} B_q = I$ for identical unit capacitances. The $q = 2$ construction is the 84-variable model for $M = 42$. Its added coordinate in each region is the local Fiedler mode, physically interpretable as the slowest internal polarization.

## S8. Multilevel convergence in the 252-node network

For the 12-region network we retained $q = 1, \ldots, 8$ local modes per region, producing 12–96 variables. The exact same three input patterns used for the 1002-node tests were applied. The point-pulse RMS errors for 12, 24, 36, 48, 60, 72, 84 and 96 variables were 6.145%, 3.959%, 2.752%, 1.387%, 0.764%, 1.181%, 0.363% and 0.242%, respectively. Uniform regional stimulation was easier: the corresponding errors were 1.161%, 0.900%, 0.527%, 0.435%, 0.385%, 0.283%, 0.098% and 0.065%.

Adjacent eigenvalue ratios in the full network ranked the gap after mode 12 first (9.419), after mode 4 second (2.010), and after mode 36 third (1.367). The 36-dimensional level corresponds to the common mode plus two internal modes per region. The weaker gaps after modes 56 and 48 had ratios 1.192 and 1.092.

The nonmonotonic error at 72 variables results from adding the sixth locally ranked mode to every region. Galerkin approximation improves in the associated variational norm, but the reported error is a particular regional output norm integrated over a finite interval. A globally ranked or balanced basis can avoid spending variables on poorly controllable or poorly observable local patterns.

## S9. Reproducibility and numerical files

The calculations were implemented in Python using NumPy and SciPy dense or sparse symmetric eigensolvers. The 1002-node robustness calculation used sparse extraction of the first 44 eigenpairs. The response comparisons used complete symmetric eigendecomposition and the analytical kernel in Eq. S5. Geometry and data visualization used Matplotlib. Random seeds are fixed in the robustness scripts. Numerical JSON files retain raw realizations and summary statistics; the 42-node resistor list reports every effective interregional conductance and resistance. The generality test of Section S11 was implemented in random_regular_ensemble.py, which builds the random-regular internal and meta-graphs directly by degree-preserving double-edge swaps (no external graph library) and uses the same modal kernel (Eq. S5) and aggregation rule (Eq. S9) as the spherical networks.

All generation, eigenanalysis, perturbation and reduced-order scripts, together with their JSON result files, are deposited in an open-access Zenodo archive under an MIT licence (DOI: to be assigned upon deposit), following FAIR and open-science practice.

## S10. Component tolerances

Exact diagonalization by the Laplacian eigenvectors requires identical capacitances and inductances. To quantify this limitation, independent lognormal variations were applied to every $C_i$ and $L_i$ in the 252-node network. For each dispersion, ten realizations were propagated with the full first-order state matrix. The reduced circuit was constructed by aggregating the actual, disordered component values rather than the nominal values.

For a point-initialized northern node, the nominal 12-region error over $0 \leq t/\sqrt{LC} \leq 45$ is 6.13%. A 1% component dispersion gives $6.22 \pm 0.28$%, a 5% dispersion gives $9.66 \pm 4.39$%, and a 10% dispersion gives $14.52 \pm 6.73$%. Thus the quotient remains useful under modest tolerances, but the exact modal interpretation progressively degrades when the local resonators cease to be identical.

## S11. Non-embedded random-regular modular ensemble

To determine which part of the hierarchy requires spherical geometry, twenty independent non-embedded networks were generated. Each network contains 24 regions of 35–50 nodes. Every region is an independently randomized 10-regular graph, and the regions are connected by a randomized 6-regular meta-graph with four weak physical edges per meta-link. Internal conductances are unity and boundary conductances are $r = 0.03$. No spatial coordinates or manifold construction enter the calculation.

Across the ensemble, the regional gap is $g_{24} = 162.5 \pm 6.9$ (range 149.9–175.1), and the mean regional character is $0.999987 \pm 0.000001$. These numbers should not be compared quantitatively with the spherical values because internal degree and boundary density are different. Their significance is qualitative: strong internal mixing and weak external coupling generate a regional band without geometric embedding. Conversely, the largest quotient-level eigenvalue ratio is only $1.28 \pm 0.14$ and its position varies between realizations. The specific four-mode global level is therefore associated with the approximate spherical geometry, not with modularity alone.

## S12. Physical status of enriched modal reductions

For a local basis $B_q$ containing $q > 1$ modes per region, the projected coupling matrix is

$$\mathcal{L}_q = B_q^T \mathcal{L} B_q.$$

It is symmetric and positive semidefinite, so the Galerkin dynamics is passive. However, the off-diagonal entries do not all have the non-positive sign required of a graph Laplacian. In the 252-node network the 24-, 36- and 96-variable matrices contain respectively 110, 280 and 2188 positive off-diagonal entries above $10^{-11}$. Consequently, these enriched models are not ordinary resistor graphs when their modal coordinates are treated as node voltages. They are passive reduced multiports and may be implemented numerically or synthesized with a more general passive network. This does not affect the direct constructibility of the one-voltage-per-region quotient.

## S13. Frequency-domain control

For harmonic current $I(\omega)$, the nodal voltage is obtained from

$$\left[i\omega C + G_0 + G_c\mathcal{L} + \frac{1}{i\omega L}\right] v(\omega) = I(\omega).$$

The same aggregation rule gives the reduced regional admittance. A point current was applied in one region and observed in a region at graph distance three. The regional quotient reproduces the location and overall form of the resonant response, but the relative $L^2$ error of this exceptionally weak far-region transfer magnitude is 56.5%. This deliberately stringent control confirms the conclusion from the time domain: a spectral gap supports a collective subspace but does not guarantee uniform relative accuracy for a nearly cancelled transfer observable.

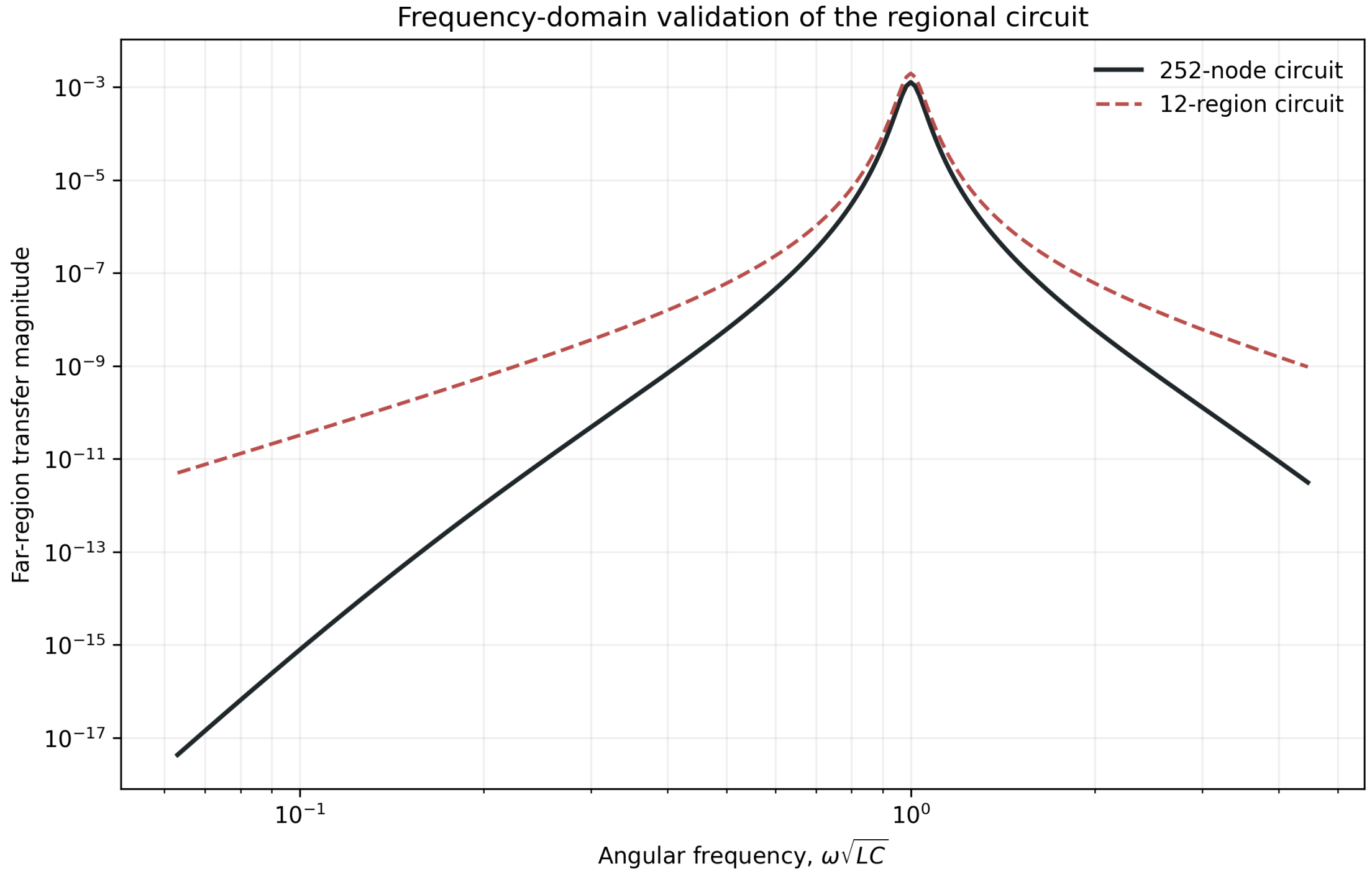

**Figure S5.** Frequency-domain response of the complete 252-node network and the directly aggregated 12-region circuit for a point-current input and a topologically distant regional output.